\documentstyle[epsfig, subfigure]{mn}

\def\ltapprox{\raise 2pt \hbox {$<$} \kern-1.1em \lower 5pt \hbox {$\approx$}}
\def\ltsim{\raise 2pt \hbox {$<$} \kern-1.1em \lower 4pt \hbox {$\sim$}}
\def\gtsim{\raise 2pt \hbox {$>$} \kern-1.1em \lower 4pt \hbox {$\sim$}}

\title{Acceleration of primary and secondary particles 
in galaxy clusters by compressible MHD turbulence : from radio halos to
gamma rays}

\author[G. Brunetti, A. Lazarian]
      {G. Brunetti,$^1$ 
       A. Lazarian$^2$\\
       $^1$ INAF/Istituto di Radioastronomia, via Gobetti 101,
       I--40129 Bologna, Italy \\
       $^2$ Department of Astronomy, University of Wisconsin at Madison,
       5534 Sterling Hall, 475 North Charter Street, Madison,
       WI 53706, USA\\
}

\begin{document}
\maketitle

\begin{abstract}
Radio observations discovered large scale non thermal sources in the 
central Mpc regions of dynamically disturbed galaxy clusters (radio halos).
The morphological and spectral properties of these sources suggest that 
the emitting electrons are accelerated by spatially distributed and gentle 
mechanisms, providing some indirect evidence for turbulent acceleration in the 
inter-galactic-medium (IGM).

\noindent
Only deep upper limits to the energy associated with relativistic protons in 
the IGM have been recently obtained through gamma and radio observations.
Yet these protons should be (theoretically) the main non-thermal particle component 
in the IGM implying the unavoidable production, at some level, 
of secondary particles that may have 
a deep impact on the gamma ray and radio properties of galaxy clysters.

Following Brunetti \& Lazarian (2007), in this paper we consider the advances in the
theory of MHD turbulence to develop a comprehensive picture of
turbulence in the IGM and extend our previous calculations of particle acceleration by 
compressible MHD turbulence by considering self-consistently the 
reacceleration of both primary and secondary particles. 
Under these conditions we expect that radio to gamma ray emission 
is generated from galaxy clusters with a complex spectrum that depends on  
the dynamics of the thermal gas and Dark Matter.
The non-thermal emission results in very good agreement with radio observations 
and with present constraints from hard X-ray and gamma ray observations. 
In our model giant radio halos are generated in mering (turbulent) clusters only. 
However, in case secondaries dominate 
the electron component in the IGM, we expect that the level of the Mpc-scale
synchrotron emission in more relaxed clusters is already close to that of the radio 
upper limits derived by present observations of clusters without radio halos.
Important constraints on cluster physics from future observations with
present and future telescopes are also discussed.
\end{abstract}

\begin{keywords}
acceleration of particles - turbulence - 
radiation mechanisms: non--thermal -
galaxies: clusters: general -
radio continuum: general - X--rays: general
\end{keywords}

\maketitle

\section{Introduction}

Radio observations of galaxy clusters prove the presence of 
non-thermal components, magnetic fields and relativistic particles, 
mixed with the hot Inter-Galactic-Medium (IGM) (e.g. Ferrari et al, 2008).

Potentially, cluster mergers can be responsible for the origin of the
non-thermal components in IGM.
During these events a fraction of the gravitational binding--energy 
of Dark Matter halos that is converted into internal energy of the barionic matter
can be channelled into the amplification of the magnetic fields
(e.g. Dolag et al. 2002; Subramanian et al. 2006; Ryu et al. 2008) and into
the acceleration of particles via shocks and turbulence (e.g. Ensslin et al 1998;
Sarazin 1999; 
Blasi 2001; Brunetti et al. 2001, 2004; Petrosian 2001; Miniati et al. 2001; 
Fujita et al 2003; Gabici \& Blasi 2003;
Ryu et al. 2003; Cassano \& Brunetti 2005; Hoeft \& Bruggen 2007; 
Brunetti \& Lazarian 2007; Pfrommer 2008).

\noindent
Theoretically relativistic protons are expected to be the dominant
non-thermal particles component since they have long life-times and
remain confined within galaxy clusters
for a Hubble time (V\"{o}lk et al. 1996; Berezinsky, Blasi \& Ptuskin 1997;
Ensslin et al 1998).
Proton--proton collisions in the IGM inject secondary particles, including
neutral pions that decay into gamma rays and relativistic electrons that
produce synchrotron and inverse Compton (IC) emission.
So far only upper limits to the gamma ray emission from galaxy
clusters have been obtained by FERMI and Cherenkov telescopes
(eg., Aharonian et al 2009a,b; Aleksic et al 2010; Ackermann et al 2010). 
These upper limits, together with constraints from complementary approaches 
based on radio observations (eg., Reimer et al 2004; Brunetti et al. 2007)
suggest that relativistic protons contribute to less than a few percent
of the energy of the IGM, at least in the central Mpc--sized regions.

Relativistic electrons in the IGM are nowadays studied by radio 
observations of diffuse synchrotron radiation from galaxy clusters.
Radio halos are the most spectacular examples of cluster-scale radio sources, 
they are diffuse radio sources that extend on
Mpc-scales in the cluster central regions and are found in about $1/3$ of 
massive galaxy clusters (eg. Feretti 2002; Ferrari et al. 2008; Cassano 2009).

\noindent 
The origin of relativistic electrons in radio halos is still debated.
In the context of the {\it hadronic} model (Dennison 1980;
Blasi \& Colafrancesco 1999; Pfrommer \& Ensslin 2004) radio halos are due to 
synchrotron emission from secondary electrons generated by p-p collisions, in
which case clusters are (unavoidably) gamma ray emitters due to the decay of
the $\pi^o$ produced by the same collisions.
The very recent non-detections of nearby galaxy clusters at GeV energies by FERMI
significantly constrain the role of secondary electrons in the non-thermal emission 
(Ackermann et al 2010). Similarly, previous works pointed out that
the spectral and morphological properties of a 
number of radio halos appear inconsistent with a simple hadronic origin of the emitting
particles (eg. Brunetti et al 2008, 2009; Donnert et al 2010a,b; 
Macario et al 2010).

\noindent
A second scenario proposed for the origin of radio halos is based
on turbulent reacceleration of relativistic particles in 
connection with cluster--mergers events (eg.,
Brunetti et al. 2001; Petrosian 2001; Berrington \& Dermer 2003;
Fujita et al 2003; Cassano \& Brunetti 2005).
The acceleration of thermal electrons to relativistic energies by 
MHD turbulence in the IGM faces serious drawbacks based on energy 
arguments (eg., Petrosian \& East 2008), consequently in these models 
it must be assumed a pre-existing population of relativistic electrons in
the cluster volume that provides the seed particles to reaccelerate during
mergers.
These seeds may be primary electrons injected by SN, AGN, galaxies and
shocks in the cluster volume that can be accumulated for a few Gyrs at energies
of a few hundred MeV (eg., Sarazin 1999; Brunetti et al 2001).

MHD turbulence theory seriously advanced in the last decades also affecting 
our view of particle acceleration in astrophysical plasmas (e.g. Chandran 2000;
Yan \& Lazarian 2002; Lazarian 2006a and ref therein).
For this reason in Brunetti \& Lazarian (2007) we considered the advances in the
theory of MHD turbulence to develop a comprehensive picture of
turbulence in the IGM and to study the reacceleration of relativistic particles.
In this respect, our main conclusions were that the compressible
MHD turbulence (essentially fact modes), generated in
connection with energetic cluster--mergers, is the most important source of stochastic
particle reacceleration in the IGM and that the interaction between this
turbulence and the relativistic electrons may explain the origin of radio halos, 
provided that enough seed relativistic electrons are accumulated in the IGM.

\noindent
At the same time relativistic protons should be accumulated (at some level) in the IGM 
and consequently MHD turbulence may also reaccelerate these protons and their
secondary products (Brunetti \& Blasi 2005).
On one hand, in this case secondaries come into the picture as a natural
pool of seed particles, on the other hand the reacceleration of relativistic
protons is expected to enhance the production rate of these secondaries with the
possible drawback that this would overcome the effect of reacceleration, thus making
the overall picture unsustainable.

\noindent
Previous papers addressed this problem and model turbulent reacceleration of both primary
and secondary particles in the IGM in the context of cluster--mergers driven
turbulence (Brunetti \& Blasi 2005; Brunetti et al 2009a).
These papers explored the complex non--thermal spectrum from galaxy clusters that is  
expected in this situation and first predicted the interplay of a long--living
spectral component, due to relativistic protons and their chain of secondary
particles, with a transient emission, leading to the formation of radio halos,
that is due to the reacceleration of electrons in cluster mergers. 
Remarkably, in this general situation gamma ray emission 
is expected from galaxy clusters, although at a lower level than that expected from 
the pure hadronic case (eg. Brunetti et al 2009a)\footnote{assuming the same
cluster magnetic field and for a given synchrotron luminosity}.

\noindent
All these papers addressing the reacceleration of primary and secondary particles
focus on Alfv\'enic reacceleration and assume the injection of Alfv\'en modes 
at small, resonant, scales, in which case however it is difficult to derive a overall 
self-consistent picture connecting clusters mergers and the generation of these modes 
at such small scales. 
This makes necessary an exploration of this picture in the context of a more 
comprehensive model of MHD turbulence in clusters, that is the goal of our paper.

\noindent
Our study is also timely due to the recent gamma and radio observations 
that put severe constraints to the energy
density of the relativistic proton component in galaxy clusters 
(Brunetti et al 2007; Aharonian et al 2009a,b; Aleksic et al 2010; Ackermann
et al 2010), and that allows
for including this component in models with substantially less degree of
freedom than in the past.

\noindent
For these reasons, in this paper we extend our calculations presented
in Brunetti \& Lazarian (2007) by considering self-consistently relativistic
protons, the generation of
secondary particles in the IGM and their reacceleration by MHD turbulence generated
in cluster mergers.

In Section 2 we discuss the properties of turbulence in galaxy
clusters, in Sect.3 we summarize the formalism for particle acceleration and
evolution, in Sect.4 we first discuss the connection between mergers and particle
reacceleration by turbulence and then show our results,
and in Sect.5 we give our conclusions and discuss model simplifications, 
future extensions of the work and a comparison with 
previous works.

\noindent
A $\Lambda$CDM cosmology with $H_0=70$ km$/$s$/$Mpc, $\Lambda=0.7$ and $\Omega_0=
0.3$ is assumed.

\section{Turbulence in the IGM}

Numerical simulations of galaxy clusters
suggest that turbulent motions may store an appreciable
fraction, 5--30\%, of the thermal energy of the IGM (e.g., Roettiger, Burns, Loken
1996; Roettiger, Loken, Burns 1997; Ricker \& Sarazin 2001; Sunyaev, Bryan \& Norman 2003; 
Dolag et al.~2005; Vazza et al.~2006, 2009a; Paul et al 2010).
The largest turbulent eddies should decay into a turbulent velocity field
on smaller scales, possibly developing a turbulent cascade. 

It is known that the mean free path of thermal protons arising from
Coulomb collisions in the hot IGM may be very large, ten to hundred kpc.
Fluids in such a collisionless regime can be very different from their collisional 
counterparts (Schekochihin et al. 2005; 2010). 
The parallel to magnetic field viscosity of IGM can
be very large and the collisionless plasmas are subject to various instabilities.
Those, however, we believe change the effective collisionality of the fluid, 
justifying the application of MHD to describing the IGM at least on its large 
scales (see Lazarian et al. 2010).
Indeed, particles in plasmas can interact through the mediation of the perturbed
magnetic fields, and effective collisionality of plasmas may differ dramatically from
the textbook estimates. The difference stems from the very instabilities that
are present in the IGM plasmas (firehose, mirror, gyroresonance etc.). 
These instabilities are expected to 
transfer the energy from the turbulent compressions on the
scales less or equal to the particle mean free path to the perturbations at the
particle gyroscale. This has been described in Lazarian \& Beresnyak (2006), 
as a result of the scattering, the mean free path of particles
decreases, which makes the fluid essentially collisional over a wide range of
scales, with the critical scale for which the fluid gets effectively collisional 
that is expected to decrease with the increase of turbulent driving rate. 
While some parts of the aforementioned paper dealing with the interaction of
turbulence and cosmic rays remain controversial, a similar approach should be
reliably applicable to thermal plasma particles. 
Therefore we believe that on scales
much larger that the thermal particle gyroradius the turbulence can be treated in
the MHD approximation 
and shall apply below the theory of MHD turbulence to the IGM.  

Turbulence generated during cluster mergers is expected to be
injected at large scales, $L_o \sim 100-400$ kpc, with 
typical velocity of the turbulent eddies at the injection scale
around $V_o \sim 300-700$ km/s (eg. Subramanian et al 2006).
This makes turbulence sub--sonic,
with $M_s = V_o/c_s \approx 0.25-0.6$, but
strongly super-Alfv\'enic, with $M_A=V_o/v_A \approx 5-10$.
Turbulent motions at large scales are thus essentially hydrodynamics and
the cascading of compressive (magnetosonic) modes may couple with
that of solenoidal motions (Kolmogorov eddies).

In Brunetti \& Lazarian (2007) we discussed that the
important consequence of the turbulence in magnetized plasma is that
both solenoidal and compressive modes in hot galaxy clusters
would not be strongly affected by viscosity at large scales
and an inertial range 
is established, provided that the velocity of the eddies at large
scales exceeds $\approx$ 300 km/s.

\noindent
In the Kolmogorov cascade the turbulent velocity $V_l$ scales as 
$V_L (l/L_o)^{1/3}$, and at scales less than $l_A\sim L_o M^{-3}_A$ 
the turbulence gets sub-Alfvenic and we enter into the MHD
regime (see discussion in Lazarian 2006b). 
For the parameters above the scale $l_A \sim 0.1-1$ kpc, but the actual 
number of $M_A$ is not certain and therefore our estimate of $l_A$ should 
be treated with caution. 
Fortunately, the above uncertainty does not change the results of our 
paper appreciably. 

Compressible MHD turbulence is a subject where a number of important 
insights have been obtained much before these ideas can be tested; 
the pioneering works in the area include Montgomery \& Turner (1981),  
Shebalin et al. (1983), Higdon (1984). 
The key idea of critical balance by Goldreich \& Sridhar (1995) has 
influenced in a profound way our further thinking of the MHD cascade.   

\noindent
In the MHD regime, at smaller scales where $V_l \leq v_A$,  
three types of modes should exist in a compressible
magnetized plasma: Alfv\'en, slow and fast modes.
Slow and fast modes may be roughly thought as the MHD counterpart of
the compressible modes, while Alfv\'en modes may be thought
as the MHD counterpart of solenoidal Kolmogorov eddies
(a more extended discussion can be found in 
Cho, Lazarian \& Vishniac 2002, and ref. therein).

Turbulence in the IGM is most likely a complex mixture
of several turbulent modes.
We shall assume that a sizeable part of turbulence
at large scales (namely at scales where the magnetic tension does not
affect the turbulent motions) is in the form of compressible motions.
This is reasonable as these modes are easily generated in high beta
medium (eg. Brunetti \& Lazarian 2007 and ref. therein).

\noindent
Situation may be radically different at smaller scales
where the magnetic field tension affects turbulent motions, i.e.
in the MHD-- regime, $l \leq l_A$.
In this case, MHD numerical simulations have shown that
a solenoidal turbulent forcing gets
the ratio between the amplitude of Alfv\'en and 
fast modes in the form (Cho \& Lazarian 2003) :

\begin{equation}
{{(\delta V)_c^2}\over{(\delta V)_s^2}}
\sim
{{(\delta V)_s v_A }\over{ c_s^2 + v_A^2 }}
\label{transfercl03}
\end{equation}

\noindent
which essentially means that coupling between these two
modes may be important only at $l \approx l_A$ (in the 
MHD-- regime it should be $(\delta V)_s \leq v_A$)
since the drain of energy
from Alfv\'enic cascade is marginal when the
amplitudes of perturbations become weaker.
Most importantly in galaxy clusters it is $c_s^2 >> v_A^2$
and thus the ratio between the amplitude of 
Alfv\'en and fast modes at scales $l < l_A$ 
is expected to be small, 
$(\delta V)_c^2/(\delta V)_s^2 \leq 
(v_A / c_s)^2 \sim 10^{-2}$ (this for solenoidal forcing at $l \approx l_A$).

\noindent
A more recent work by Kowal \& Lazarian (2010) decomposed turbulent 
motions into slow, fast and Alfven modes using wavelets. 
This approach is better justified than the decomposition in Fourier space 
employed in Cho \& Lazarian (2002, 2003). 
Indeed, Alfven and slow modes are defined in the local system of coordinates 
(see Lazarian \& Vishniac 1999, Cho \& Vishniac 2000, Maron \& 
Goldreich 2001) and therefore the procedure of Fourier decomposition 
by Cho \& Lazarian (2002, 2003, see 2005 for a review) can only be 
statistically true. The wavelet decomposition procedure is more localized 
in space and therefore is potentially more precise. Nevertheless, the 
results on mode decomposition in Kowal \& Lazarian (2010) 
agree well with those in Cho \& Lazarian (2003), which 
provides us with more confidence about the properties of MHD turbulence 
that we employ in the paper to describe turbulence interaction with 
energetic particles.

Our treatment of turbulence assumes that the turbulence is balanced, i.e. 
the energy flux of waves\footnote{Strong MHD turbulence presents a case 
of dualism of waves and eddies (see Lazarian \& Vishniac 1999; 
Cho, Lazarian \& Vishniac 2003). Strong non-linear damping of oppositely 
moving Alfvenic wave packets makes them act like eddies.} moving in one 
direction is equal to the energy flux in the opposite direction. 
While the theory of imbalanced turbulence is being intensively developed 
(see Lithwick \& Goldreich 2001; Beresnyak \& Lazarian 2008, 2009, 2010;
Chandran 2008, Perez \& Boldyrev 2009) we do not expect that the effects of
imbalance would dominate in the cluster environments. 
First of all, the turbulence driving is not expected to be strongly 
localized and then the effects of compressibility should decrease the 
local turbulence imbalance.

\section{Particle acceleration, energy losses and secondary particles}

In this paper we provide an extension of our previous calculations.
We assume the picture of MHD turbulence in the IGM, as derived
in Brunetti \& Lazarian (2007), to calculate the reacceleration of relativistic
particles by compressible MHD turbulence by taking into account self-consistently 
also the generation and reacceleration of secondary particles.

\noindent 
The aim of this Section is to present the formalism and the main 
assumptions used in our calculations.

\subsection{Basic formalism}

We model the re-acceleration of
relativistic particles by MHD turbulence in the most simple
situation in which only relativistic protons
are initially present in a turbulent IGM. These protons generate
secondary electrons via p-p collisions and in turns secondaries (as
well as protons) are reaccelerated by MHD turbulence. 

We model the time evolution of the spectral energy 
distribution of electrons, $N_e^-$, and positrons, $N_e^+$,
with an isotropic Fokker-Planck
equation\footnote{isotropization of particles is discussed in Sect.~3.2.} :

\begin{eqnarray}
{{\partial N_e^{\pm}(p,t)}\over{\partial t}}=
{{\partial }\over{\partial p}}
\Big[
N_e^{\pm}(p,t)\Big(
\left|{{dp}\over{dt}}_{\rm r}\right| -
{1\over{p^2}}{{\partial }\over{\partial p}}(p^2 D_{\rm pp}^{\pm})
\nonumber\\
+ \left|{{dp}\over{dt}}_{\rm i}
\right| \Big)\Big]
+ {{\partial^2 }\over{\partial p^2}}
\left[
D_{\rm pp}^{\pm} N_e^{\pm}(p,t) \right] \nonumber\\
+ Q_e^{\pm}[p,t;N_p(p,t)] \, ,
\label{elettroni}
\end{eqnarray}

\noindent
where $|dp/dt|$ marks radiative (r) and Coulomb (i) losses (Sect.~3.3),
$D_{pp}$ is the electron/positron diffusion coefficient in the momentum
space due to the coupling with magnetosonic modes (Sect.~3.2), 
and the term $Q_{e}^{\pm}$ accounts for the injection rate of secondary 
electrons and positrons due to p-p collisions in the IGM (Sect.~3.4).

\noindent
The time evolution of the spectral energy
distribution of protons, $N_p$, is given by :

\begin{eqnarray}
{{\partial N_p(p,t)}\over{\partial t}}=
{{\partial }\over{\partial p}}
\Big[
N_p(p,t)\Big( \left|{{dp}\over{dt}}_{\rm i}\right|
-{1\over{p^2}}{{\partial }\over{\partial p}}(p^2 D_{\rm pp})
\Big)\Big]
\nonumber\\
+ {{\partial^2 }\over{\partial p^2}}
\left[ D_{\rm pp} N_p(p,t) \right] - {{N_p(p,t)}\over{\tau_{pp}(p)}}
\, ,
\label{protoni}
\end{eqnarray}

\noindent
where $|dp/dt_i|$ marks Coulomb losses (Sect.~3.3),
$D_{pp}$ is the diffusion coefficient in the momentum
space of protons due to the coupling with magnetosonic
modes (Sect.~3.2), and $\tau_{pp}$ is the proton life--time 
due to pp collisions in the IGM (Sect.~3.3; see also Ensslin et al 2007).

\noindent
For isotropic turbulence (Sect.~3.2) the diffusion equation in the
k--space is given by :

\begin{eqnarray}
{{\partial {\cal W}(k,t) }\over{\partial t}}
=
{{\partial}\over{\partial k}}
\left(
k^2 D_{kk}
{{\partial}\over{\partial k}}
( {{ {\cal W}(k,t) }\over{k^2}} )
\right) + I(k,t) \nonumber\\
- \sum_i \Gamma_i (k,t) {\cal W}(k,t) \,\,\, ,
\label{modes_kinetic}
\end{eqnarray}

\noindent
where $D_{kk}$ is the diffusion coefficient in the k--space,
$\Gamma_i(k,t)$ are the different damping terms,
and $I(k,t) = I_o \delta(k-k_o)$ is the turbulence injection term, i.e.
we consider the most simple situation where turbulence is injected
at a single scale, with wavenumber $k_o$.
The wave--wave diffusion coefficient of magnetosonic modes
(Kraichnan treatment) is given by (Brunetti \& Lazarian 2007 and ref
therein): 

\begin{equation}
D_{kk}
\approx \langle V_{\rm ph} \rangle k^4 
\left(
{{ {\cal W}(k,t) }\over{\rho \langle V_{\rm ph} \rangle^2 }}
\right)
\label{Dkk}
\end{equation}

\noindent
where $\langle V_{\rm ph} \rangle$ is a representative, averaged (with
respect to $\theta$), phase velocity.

We shall assume isotropic MHD turbulence 
to calculate the particle acceleration
rate at any time.
This assumption is appropriate for super--Alfv\'enic turbulence and
fast modes (e.g., Cho \& Lazarian 2003), provided that collisionless
dampings are not efficient.
At smaller scales, collisionless dampings with thermal particles in the
IGM become severe and modify the spectrum of turbulent modes, and
a cut-off in the turbulent spectrum is generated at $k = k_c$, where
the damping time--scale becomes shorter than the cascading time.
The most important damping with thermal (and relativistic) particles
is the Transit-Time-Damping (TTD)
that is highly anisotropic (e.g. Schlickeiser \& Miller 1998;
Brunetti \& Lazarian 2007; Yan et al. 2008) being stronger for 
$\theta \sim \pi/2$ and causing the spectrum of the turbulent modes to
become anisotropic at scales $k \sim k_c (\theta)$.
On the other hand, in Brunetti \& Lazarian (2007) we
have shown that, under physical conditions typical of the IGM, 
hydro--motions bend the magnetic-field lines in a time--scale,
$\tau_{bb} \approx l_A/v_A$, that is comparable to the damping time--scale
of turbulence, in which case there is chance that
approximate isotropization of the turbulent spectrum is maintained
even at scales where dampings are severe.

\noindent
Thus following Brunetti \& Lazarian (2007) we shall assume a simplified
turbulent (isotropic) spectrum in the form :

\begin{equation}
{\cal W}(k) \approx 
\left(
I_o \rho \langle V_{\rm ph}\rangle \right)^{ {1\over 2} }
k^{- {3 \over 2} }
\label{cascade_spectrum}
\end{equation}

\noindent
for $k_o < k < k_{c}$,
with the cut-off wavenumber estimated from the condition that the damping time-scale
becomes smaller than the cascading time scale :

\begin{equation}
\tau_d \approx 1 / {{\langle \sum \Gamma_i(k,\theta) \rangle}}
= \tau_{kk} \approx k^2/D_{kk}
\end{equation}

\noindent
that is :

\begin{equation}
k_{c} = {\cal C}
{{I_o}\over{\rho \langle V_{\rm ph} \rangle}}
\left( {{\langle \sum \Gamma_i(k,\theta) \rangle}\over{k}} \right)^{-2}
\label{kc}
\end{equation}

\noindent
where quantities $\langle .. \rangle$ are averaged with respect to $\theta$
and $\sum \Gamma_i$ is the total (thermal and non-thermal) damping
term due to TTD resonance and the constant ${\cal C} \sim$ a few 
(Brunetti \& Lazarian 2007; see also Matthaeus \& Zhou 1989,
for details on Kraichnan constants).

\subsection{Momentum diffusion coefficient due to compressible
MHD turbulence}

Compressible turbulence can affect particle  
motion through the action of the
mode--electric field via gyroresonant interaction (e.g.,
Melrose 1968), the condition for which is :

\begin{equation}
\omega
- k_{\Vert}v_{\Vert}
-n {{\Omega}\over{\gamma}} =0
\label{resonance}
\end{equation}

\noindent
where $n=\pm 1$, $\pm 2$, .. gives the
first (fundamental), second, .. harmonics
of the resonance, while
$v_{\Vert}=\mu v$ and $k_{\Vert}=\eta k$
are the parallel (projected along the magnetic field)
speed of the
particles and the wave--number, respectively.

Following Brunetti \& Lazarian (2007) we assume 
particle--mode coupling through the 
Transit--Time Damping (TTD), $n=0$, resonance (e.g., Fisk 1976; Eilek 1979;
Miller, Larosa \& Moore 1996; Schlickeiser \& Miller 1998).
In principle, this resonance changes only the
component of the particle momentum parallel to the
seed magnetic field and this would cause an increasing degree of anisotropy
of the particle distribution leading to a less and less
efficient process with time.
Thus an important aspect in this working picture is the need of
isotropization of particle momenta during acceleration
(e.g., Schlickeiser \& Miller 1998).
In this paper we shall assume continuous isotropization of particle
momenta. 
Isotropy may be provided by several processes discussed in the literature. 
These include electron firehose instability 
(Pilipp \& V\"olk 1971; Paesold \& Benz 1999), and 
gyro-resonance by Alfv\'en (and slow) modes at small scales, provided
that these modes are not too much anisotropic (cf. Yan \& Lazarian
2004)\footnote{The latter condition means that the Alfv\'enic modes are
considered for scales not much less than $l_A$, provided that the
turbulence injection is isotropic}. Gyro-resonance may also
occur with electrostatic lower hybrid modes generated by
anomalous Doppler resonance instability due to
pitch angle anisotropies
(e.g., Liu \& Mok 1977; Moghaddam--Taaheri et al.
1985) and, possibly, with whistlers (e.g., Steinacker \& Miller 1992). 
In addition, Lazarian \& Beresnyak (2006) proposed isotropization of 
cosmic rays due to {\it gyroresonance 
instability} that arises as the distribution of cosmic rays gets
anisotropic in phase space. 

We adopt the momentum--diffusion coefficient of particles, $D_{pp}$, 
as derived from detailed balancing argument, i.e. relating the diffusion 
coefficient of a $\alpha$--species to the damping rate of the modes 
themselves with the same particles (eg., Eilek 1979; 
Brunetti \& Lazarian 2007)\footnote{This is obtained by assuming that 
particles isotropy is {\it maintained} during acceleration
(see Brunetti \& Lazarian 2007)} :

\begin{eqnarray}
D_{pp}(p)={{\pi^2}\over{2\, c}}
p^2
{{ 1 }\over{B_o^2}}
\int_0^{\pi/2} d\theta V_{\rm ph}^2
{{ \sin^3(\theta) }\over{ |\cos(\theta) | }}
{\cal H}(1 - {{V_{\rm ph}/c}\over{\cos \theta}} ) \nonumber\\
\times \left(
1 - ( {{V_{\rm ph}/c}\over{\cos \theta}} )^2 \right)^2
\int_{k_c} dk
{\cal W}_B(k) k
\label{dpp1}
\end{eqnarray}

\noindent
where $B_o$ is the background (unperturbed) magnetic field,
$k_c$ is given in Eq.~\ref{kc} and 

\begin{equation}
{\cal W}_B(k) =
{1\over{\beta_{pl}}}
\langle
{{\beta_{pl} |B_k|^2 }\over
{16 \pi {\cal W}(k)}}
\rangle
{\cal W}(k) \,\,\, ,
\label{wbk}
\end{equation}

\noindent
where $|B_k|^2/{\cal W}(k)$ is the ratio between magnetic field 
fluctuations and total energy in the mode (the quantity $\langle .. \rangle$
in Eq.~\ref{wbk} indicates average with respect to
$\theta$ and is of order unity, see Brunetti \& Lazarian 2007 for further
details).

\subsection{Energy losses for electrons and protons}

The energy losses of relativistic electrons in the IGM are
dominated by ionization and Coulomb losses, at low energies, 
and by synchrotron and IC losses, at higher energies (eg. Sarazin 1999). 
The rate of losses due to the 
combination of ionization and Coulomb scattering is (in cgs units):

\begin{equation}
\left( {{ d p }\over{d t}}\right)_{\rm i} 
=- 3.3 \times 10^{-29} n_{\rm th}
\left[1+ {{ {\rm ln}(\gamma/{n_{\rm th}} ) }\over{
75 }} \right]
\label{ion}
\end{equation}

\noindent
where $n_{\rm th}$ is the number density of the thermal plasma.
The rate of synchrotron and IC losses is (in cgs units):

\begin{equation}
\left( {{ d p }\over{d t}}\right)_{\rm rad}
=- 4.8 \times 10^{-4} p^2
\left[ \left( {{ B_{\mu G} }\over{
3.2}} \right)^2 {{ \sin^2\theta}\over{2/3}}
+ (1+z)^4 \right]
\label{syn+ic}
\end{equation}

\noindent
where $B_{\mu G}$ is the magnetic field strength in
units of $\mu G$, and $\theta$ is the pitch angle of the emitting 
leptons; in case of efficient isotropization of the 
electron momenta, the $\sin^2\theta$ is averaged to $2/3$.

For relativistic protons, the main channel of energy losses 
in the IGM is provided by inelastic p-p collisions.
The life--time of protons due to pp collisions is given by :

\begin{equation}
\tau_{pp}(p)=
{1\over{c \, n_{th} 
\sum \sigma^{+/-,o}}}
\label{tpp}
\end{equation}

\noindent
In this paper we use the inclusive cross section, 
$\sigma^{\pm,o}(p_p)$, given by
the fitting formulae in Dermer (1986a) which allow to describe
separately the rates of generation of neutral and charged pions.

\noindent
For trans-relativistic and mildly relativistic protons, energy
losses are dominated by ionization and Coulomb scattering. 
Protons more energetics than the thermal electrons, namely with 
$\beta_p > \beta_c \equiv (3/2 m_e/m_p)^{1/2} \beta_e$
($\beta_e \simeq 0.18 (T/10^8 K)^{1/2}$ is the velocity of the 
thermal electrons), are affected by Coulomb interactions.
Defining $x_m \equiv \left( {{ 3 \sqrt{\pi}}\over{4}}
\right)^{1/3}\beta_e$, one has (Schlickeiser, 2002):

\begin{equation}
\Big( {{ d p }\over{dt}} \Big)_i \simeq
- 1.7 \times 10^{-29}
\left( {{n_{\rm th}}\over{10^{-3}}} \right)
{{
\beta_p }\over{
x_m^3 + \beta_p^3 }} \,\,\,\,\,\,\,\,\, ({\rm cgs})
\label{coulomb_p}
\end{equation}

\subsection{Injection of Secondary Electrons}

The decay chain that we consider for the injection
of secondary particles in the IGM due to p-p
collisions is (Blasi \& Colafrancesco 1999):

$$p+p \to \pi^0 + \pi^+ + \pi^- + \rm{anything}$$
$$\pi^0 \to \gamma \gamma$$
$$\pi^\pm \to \mu + \nu_\mu ~~~ \mu^\pm\to e^\pm \nu_\mu \nu_e.$$

\noindent
that is a threshold reaction that requires protons with kinetic
energy larger than $T_p \approx 300$ MeV. 

A practical and useful approach to describe the pion spectrum
both in the high 
energy ($E_p > 10$ GeV) and low energy regimes was proposed 
in Dermer (1986b) and reviewed by Moskalenko \& Strong (1998)
and Brunetti \& Blasi (2005), and is based on the combination of the
isobaric model (Stecker 1970) and scaling model
(Badhwar et al., 1977; Stephens \& Badhwar 1981). 

\noindent 
The injection rate of pions is given by :

\begin{equation}
Q_{\pi}^{\pm,o}(E,t)= n^p_{th} c 
\int_{p_{*}} dp N_p(p,t) \beta_p {{ F_{\pi}(E_{\pi},E_p) 
\sigma^{\pm,o}(p_p)}\over
{\sqrt{1 + (m_pc/p_p)^2} }},
\label{q_pi}
\end{equation}

\noindent
where we adopted $F_{\pi}$ as given in Brunetti \& Blasi (2005)
that use the isobaric model for $E_p < 3$ GeV, a scaling model for
$E_p > 10$ GeV and a linear combination of the two models for intermediate
energies, and where $p_{tr}$ is the
threshold momentum of protons for the process to occur.

The injection rate of relativistic electrons/positrons is given by :

\begin{eqnarray}
Q_{e^{\pm}}(p,t)=
\int_{E_{\pi}}
Q_{\pi}(E_{\pi^{\pm}},t) dE_{\pi} \int dE_{\mu} \times \nonumber\\
F_{e^{\pm}}(E_{\pi},E_{\mu},E_e) F_{\mu}(E_{\mu},E_{\pi}),
\label{qepm1}
\end{eqnarray}

\noindent
where $F_e^{\pm}(E_e,E_\mu,E_\pi)$ is the spectrum of
electrons and positrons from the
decay of a muon of energy $E_\mu$ produced in the decay of a pion with
energy $E_\pi$, and $F_{\mu}(E_{\mu},E_{\pi})$ is the muon spectrum
generated by the decay of a pion of energy $E_{\pi}$ that is :

\begin{equation}
F_{\mu}(E_{\mu},E_{\pi})= 
{{ m_{\pi}^2 }\over{m_{\pi}^2- m_{\mu}^2}}
{1\over{\sqrt{E_{\pi}^2 - m_{\pi}^2}}}
\label{fmupi}
\end{equation}

\noindent
between a kinematic minimum and maximum muon energy given by :

\begin{equation}
m_{\mu} 
\gamma_{\pi} \gamma_{\mu}^{\prime}
(1 - \beta_{\pi} \beta_{\mu}^{\prime})
\leq E_{\mu} \leq m_{\mu} 
\gamma_{\pi} \gamma_{\mu}^{\prime}
(1 + \beta_{\pi} \beta_{\mu}^{\prime})
\label{emumin}
\end{equation}

where $\gamma_{\mu}^{\prime}$ is the Lorentz factor of the muon in the
pion frame, $\beta_{\mu}^{\prime} \simeq 0.2714$ (Moskalenko \& Strong 1998),
and from kinematics :

\begin{equation}
m_{\mu} \gamma_{\mu}^{\prime} \beta_{\mu}^{\prime}
=
{{m_{\pi}^2 - m_{\mu}^2 }\over
{2 m_{\pi}}} \, .
\label{emumax}
\end{equation}

\noindent
In order to simplify calculations, following Brunetti \& Blasi (2005),
we assume that the spectrum of muons is a delta--function :

\begin{equation}
F_{\mu}(E_{\mu},E_{\pi})=
\delta \left(
E_{\mu} - {\bar{E}_{\mu}}
\right)\, ,
\label{fmupi_delta}
\end{equation}

\noindent
where 

\begin{equation}
{\bar{E}_{\mu}} = 
{{m_{\pi}^2 - m_{\mu}^2 }\over
{m_{\pi}^2}} {{E_{\pi}}\over{2 \beta_{\mu}^{\prime}}}
\, .
\label{emumax}
\end{equation}

\noindent
We use the spectrum of electrons and positrons from the muon
decay, $F_{e^{\pm}}(E_{\pi},E_{\mu},E_e)$, as given by
Blasi \& Colafrancesco (1999), and combining their results with Eqs.
(\ref{q_pi}), (\ref{qepm1}) and (\ref{fmupi_delta}), 
we obtain the rate of production
of secondary electrons/positrons :

\begin{eqnarray}
Q_{e^{\pm}}(p,t) =
{{ 8 \beta_{\mu}^{\prime} m_{\pi}^2 n_{th}^p c}\over{ m_{\pi}^2 - 
m_{\mu}^2 }}
\int_{E_{\rm min}}
\int_{E_{*} } 
{{dE_{\pi} dE_p}\over{ E_{\pi} {\bar{\beta}}_{\mu} }}
\beta_p N(E_p) \times \nonumber\\
\sqrt{1 -( {{m_p c^2}\over{E_p }})^2 } 
\sigma^{\pm}(E_p) F_{\pi}(E_{\pi},E_p)\tilde{F} \, ,
\label{qepm2}
\end{eqnarray}

\noindent
where $E_{\rm min}=2 E_e \beta_{\mu}^{\prime}
m_{\pi}^2/(m_{\pi}^2 - m_{\mu}^2)$, 
and 

$$\tilde{F}= {{5}\over{12}}-{{3}\over{4}}{\bar{\lambda}}^2
+{1\over 3}{\bar{\lambda}}^3- {{ {\bar{P}}_{\mu}}
\over{2 {\bar{\beta}}_{\mu}}}
\big( {1\over 6} - ({\bar{\beta}}_{\mu} +{1\over 2}){\bar{\lambda}}^2
+({\bar{\beta}}_{\mu} + {1\over 3}){\bar{\lambda}}^3 \big),$$
$$\,\,\,\,\,\,\,\,\,\,\,\,\,\,\,\,\,\,\,\,
\,\,\,\,\,\,\,\,\,\,\,
\,\,\,\,\,\,\,\,\,\,\,\,\,\,\,\,\,\,\,\,
\,\,\,\,\,\,\,\,\,\,\,\,\,\,\,\,\,\,\,\,\,\,
{\rm for} \,\,\,\,\,\,\,\,
{{1 - {\bar{\beta}}_{\mu}}\over{1 + {\bar{\beta}}_{\mu}}} \leq
{\bar{\lambda}} \leq 1; $$
and
$$= {{ {\bar{\lambda}}^2 {\bar{\beta}}_{\mu} }\over{
(1 - {\bar{\beta}}_{\mu})^2}}\Big[
3 -{2\over 3}{\bar{\lambda}}
\left({{3 + {\bar{\beta}}_{\mu}^2 }\over{
1-{\bar{\beta}}_{\mu} }} \right) \Big] -
{{ {\bar{P}}_{\mu} }\over{1 - {\bar{\beta}}_{\mu} }}
\Big\{
{\bar{\lambda}}^2 (1+ {\bar{\beta}}_{\mu}) - $$
$$\,\,\,\,\,\,\,\,\,\,\,\,\,\,\,\,\,\,\,\,
\,\,\,\,\,\,\,\,\,\,\,\,\,\,\,\,\,\,\,\,\,\,\,\,
{{ 2 {\bar{\lambda}}^2}\over{1 - {\bar{\beta}}_{\mu}}}
\left[ {1\over 2} + {\bar{\lambda}}
(1 + {\bar{\beta}}_{\mu} ) \right]
+ {{ 2 {\bar{\lambda}}^3 (3 + {\bar{\beta}}_{\mu}^2) }\over
{ 3 (1- {\bar{\beta}}_{\mu})^2 }}
\Big\},$$
$$\,\,\,\,\,\,\,\,\,\,\,\,\,\,\,\,\,\,\,\,
\,\,\,\,\,\,\,\,\,\,\,
\,\,\,\,\,\,\,\,\,\,\,\,\,\,\,\,\,\,\,\,
\,\,\,\,\,\,\,\,\,\,\,\,\,\,\,\,\,\,\,\,\,\,
{\rm for} \,\,\,\,\,\,\,\,
0 \leq {\bar{\lambda}} \leq {{1 - {\bar{\beta}}_{\mu}}\over{1 +
{\bar{\beta}}_{\mu}}}.$$

\noindent
and where we use the following definitions :

\begin{equation}
{\bar{\lambda}}=
{{E_e}\over{\bar{E}_{\mu}}}=
{{E_e}\over{E_{\pi}}}
\left({{2 \beta_{\mu}^{\prime} m_{\pi}^2 }\over
{m_{\pi}^2 - m_{\mu}^2}}
\right) ,
\end{equation}

\begin{equation}
{\bar{\beta}}_{\mu}=
\left(1 - {{m_{\mu}^2 }\over{ {\bar{E}_{\mu}^2}}}
\right)^{1/2}
\end{equation}

\noindent
and 

\begin{equation}
{\bar{P}}_{\mu}=
- {1\over{{\bar{\beta}}_{\mu}}}
{{ m_{\pi}^4}\over{(m_{\pi}^2 - m_{\mu}^2)^2 }}
\left[ 4 \beta_{\mu}^{\prime} -1 + ( {{m_{\mu} }\over{ m_{\pi}}} )^4
\right] \, .
\label{ppi}
\end{equation}

\section{Results}

\begin{figure*}
\begin{center}
\includegraphics[width=0.34\textwidth]{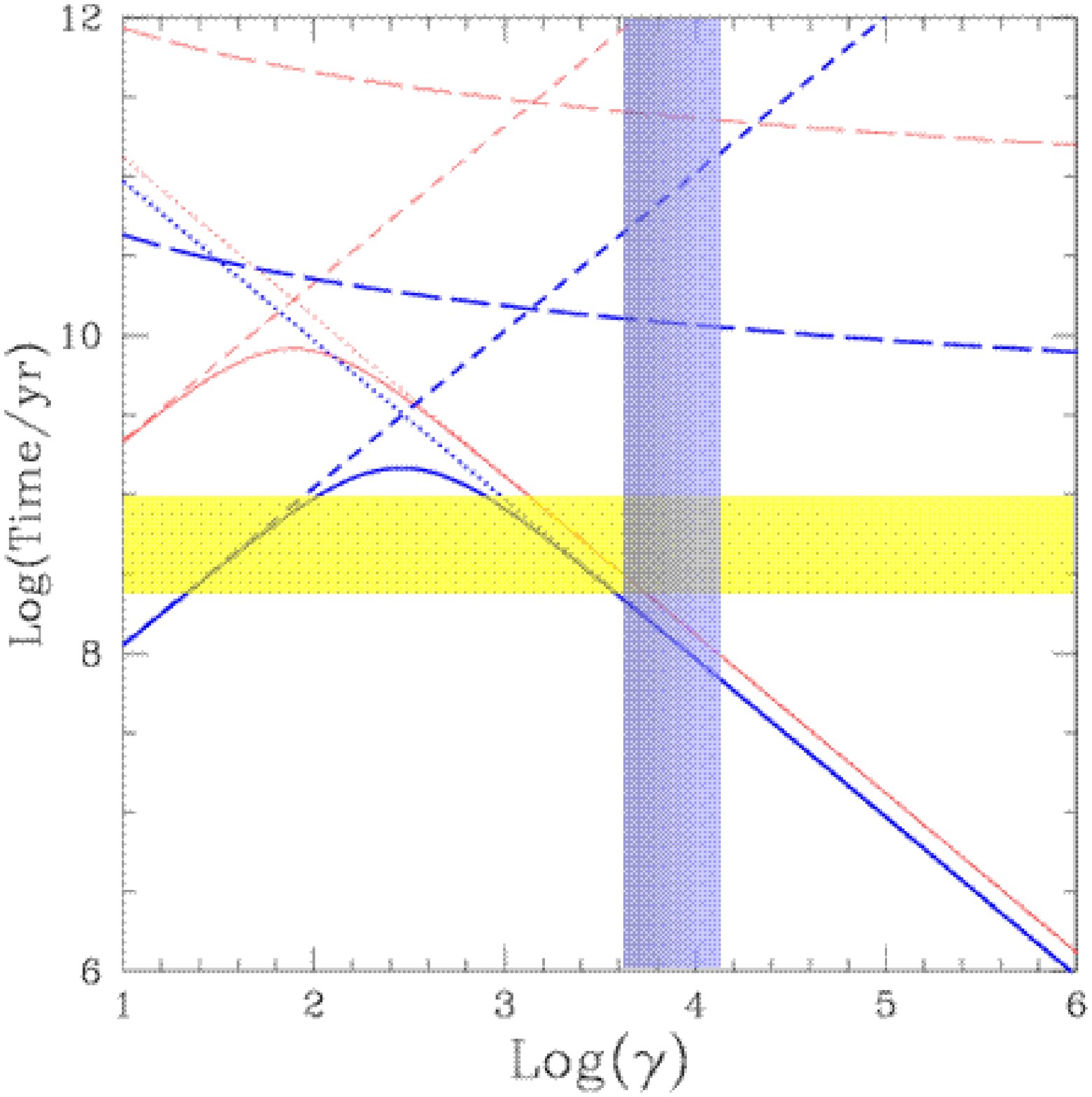}%
\includegraphics[width=0.35\textwidth]{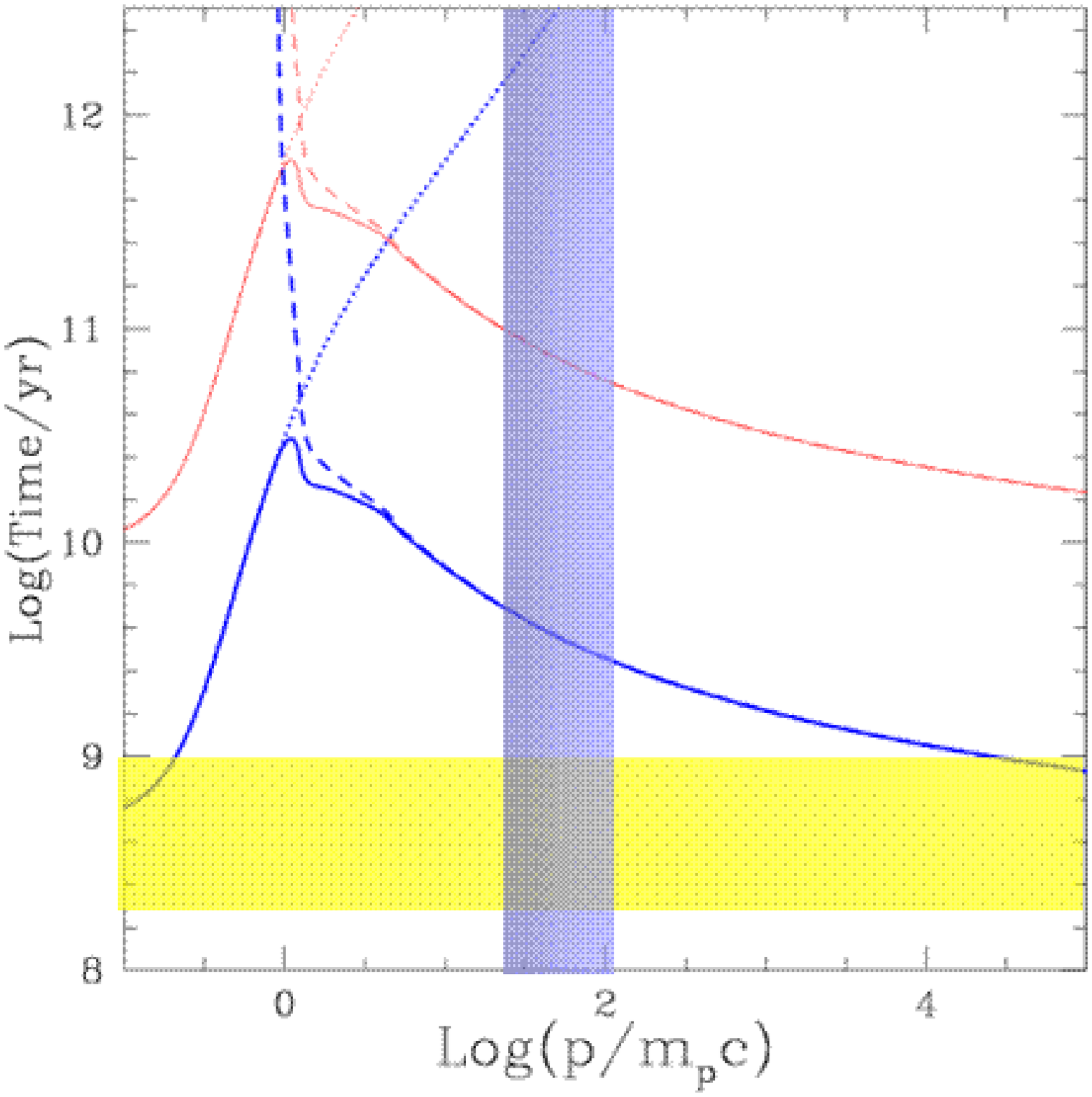}
\caption{{\bf Left panel}: The life--time of relativistic electrons in the
IGM at $z=0.2$ as a function of their Lorentz factor.
Thick (blue) lines are for cluster cores ($B=3 \mu$G, $n_{th}=2\times
10^{-3}$cm$^{-3}$) and thin (red) lines
are for cluster periphery ($B=0.5 \mu$G, $n_{th}=10^{-4}$cm$^{-3}$).
We report the total life--time (solid lines) and the life--times
due to single processes: Coulomb losses (dashed lines),
synchrotron and IC losses (dotted lines), and bremsstrahlung
losses (long dashed lines).
The yellow region marks the turbulence eddy turnover time 
(assuming 
$L_o \sim 200-300$ kpc, $(V_L/c_s)^2 \approx 0.1-0.3$, $T \approx 10^8$ K),
while the blue region marks the range of Lorentz factors of the 
relativistic electrons emitting at frequencies 
$\approx 300-1400$ MHz.
{\bf Right panel}: The life--time of cosmic--ray protons
in the IGM as a function of the particle momentum.
Thick (blue) lines are for cluster cores and thin (red) lines
are for cluster periphery.
We report the total life--time (solid lines) and the life--times
due to single processes: Coulomb losses (dotted lines) and
pp--collisions (dashed lines).
The yellow region marks the turbulence eddy turnover time, 
while the blue region marks the range of momentum of
relativistic protons that mostly contribute to the injection of
secondary electrons emitting at frequencies $\approx 300-1400$ MHz.
}
\label{fig:times_ep}
\end{center}
\end{figure*}

In this Section we calculate particle acceleration and non-thermal
emission in galaxy clusters by assuming that 
MHD turbulence is generated during mergers between clusters and reaccelerates
primary and secondary particles.

\noindent
Following previous works (eg. Brunetti \& Blasi 2005) we adopt a
simplified situation where we do not consider primary electrons 
in the IGM (see Sect.5 for discussion), and where 
the two main ingredients are (i) relativistic protons, that are believed to be 
the most important non--thermal particle components in 
the IGM (eg., Blasi et al 2007 for review), and (ii) the MHD turbulence.
On one hand, relativistic protons inject secondary particles via p--p collisions 
in the IGM that produce radiation from the radio to the gamma ray band, at the same time 
MHD turbulence may reaccelerate relativistic protons and secondary electrons 
in the IGM, generating radio
halos and leaving an imprint in the general non--thermal properties of galaxy clusters.

At variance with the aforementioned paper that focus on the Alfvenic case, following
the model of MHD turbulence in galaxy clusters by Brunetti \& Lazarian (2007),  
we assume that MHD turbulence in the IGM is in 
the form of compressible modes whose cascading from large to small scales results 
in a isotropic turbulent--spectrum. 
This allows us to readily connect the
injection of turbulence at large scales with the particle acceleration process
and to study the theoretical framework of the connection between
cluster mergers and turbulent reacceleration of relativistic particles in the IGM.
As a matter of fact calculations reported in this Section provide 
an extension of those in Brunetti \& Lazarian (2007) 
that consider reacceleration of (only) primary particles by compressible
MHD modes. 

\noindent
As already stressed, 
one of the main motivations for these new calculations comes from the 
recent gamma ray and radio observations that put severe constraints on the energy
density of relativistic protons in galaxy clusters (Brunetti et al 2007; 
Aharonian et al 2009a,b; Aleksic et al 2010; Ackermann et al 2010), allowing 
for including secondary particles in turbulent--acceleration
models with substantially less degree of freedom than in the past.

\subsection{Turbulent reacceleration, time--scales and connection with mergers}

In the framework adopted in our paper the idea is that radio halos are generated
by the reacceleration of relativistic electrons (secondaries in our specific case)
by turbulence generated during cluster mergers.

\noindent
The leading processes in the context of our scenario
are the generation, cascading and dissipation of turbulence in the
IGM, and the acceleration and cooling of relativistic particles.
These processes, and the cluster-cluster collisions themselves, have
their own time-scale and the general picture breaks out from the
interplay of all these time-scales.
Figure \ref{fig:times_ep} 
reports the time-scales of the most relevant processes 
as calculated in different, relevant, regions of the
cluster volume.

The generation and dissipation 
of turbulent motions in the IGM is not studied in 
great details, however present numerical simulations suggest 
that these motions can be generated in galaxy clusters 
for a substantial fraction of the period of cluster-cluster interaction,
a few Gyr, possibly driven by shock waves that cross the cluster 
volume and by the sloshing (and stripping) of cluster cores (e.g.,
Dolag et al. 2005; Vazza et al. 2009a; Paul et al. 2010; ZuHone et al.
2010).
Under our working picture, the compressible turbulence, after being injected 
at larger scales, decays at smaller scales where it dissipates 
through dampings with thermal and non-thermal particles 
in the IGM (Sect.~3.1).
The turbulence decay requires about one eddy turnover time,
that in the case of fast mode is 
$\tau _{kk} \sim {{ V_{ph} }\over{V_L^2}} (L / k)^{1/2}$ 
(e.g., Yan \& Lazarian 2004), implying a unavoidable delay between the first 
generation of large-scale turbulence in a given region and the beginning of 
the phase of particle reacceleration (that is mainly due
to the non linear interaction of particles with turbulent modes at smaller scales)
in the same region.
We believe however that this delay does not break the temporal connection
between mergers and particle acceleration, since the eddy turnover time of
compressible turbulence in massive (hot) clusters is 
$\tau _{kk} \approx 0.2-1$ Gyr (by assuming typical injection scales $L_o
\approx 200-300$ kpc and $(V_L/c_s)^2 \approx 0.1-0.3$), that
is smaller than the typical duration of cluster-cluster interaction. 

\noindent
Under our simplified working picture, where turbulence is (only) generated by 
energetic
cluster mergers, compressible turbulence dissipates completely in a few eddy 
turnover times, as soon as galaxy clusters becomes more relaxed.
The decrease of the efficiency of turbulent--particle acceleration and 
the suppression of non-thermal cluster--scale emission, i.e. the "dissipation"
of radio halos, 
are even faster due to 
the fact that (i) the acceleration efficiency scales non 
linearly with the turbulent spectrum\footnote{e.g., 
$\tau_{acc} \propto D_{pp} \propto W_k^2$, combining Eqs.
\ref{cascade_spectrum}, \ref{kc} and \ref{dpp1}, see also Brunetti \&
Lazarian (2007)} and (ii) the cooling time of the radio--emitting electrons is
short, $\approx 0.1$ Gyr (Fig.~\ref{fig:times_ep}).
The consequence is a tight connection between radio halos and cluster mergers,
although the picture may be more complex as the 
secondary particles, continuously injected in the IGM, should generate synchrotron 
emission at some level also in relaxed clusters (see Sect. 4.3.2).

\noindent
Radio observations of statistical samples of galaxy clusters
show that (i) giant radio halos form (only) in merging clusters, (ii) that 
their life--time in merging clusters is of the order of 1 Gyr, 
and (iii) suggest that halo emission 
should dissipate in relaxed clusters in a short, $<$ Gyr, 
time-scale (e.g. Hwang 2004; Brunetti et
al. 2007, 2009b; Venturi et al 2008); these observational points are 
consistent with the working picture described in this Section.

\begin{figure}
\includegraphics[width=\columnwidth]{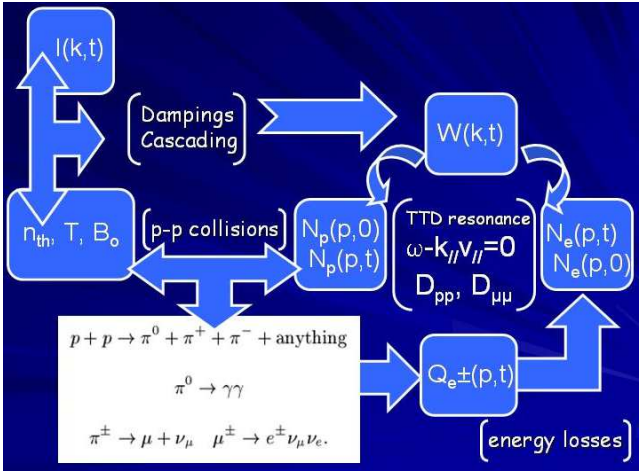}%
\caption{A scheme of the processes taken into account in our calculations and 
the coupling between them. The starting points are the properties of the
magnetized IGM ($n_{th}$, $T$, $B_o$), the initial spectrum of relativistic
protons ($N_p(p,0)$) and the injection rate of turbulence ($I(k)$).}
\label{fig:schema}
\end{figure}

\subsection{Spectral evolution of reaccelerated
protons and electrons}

In this Section we report on some relevant results on 
the evolution of the particles spectrum (protons
and secondary electrons/positrons) subject to TTD resonance
with compressible MHD turbulence; the approach followed in this section is
than used in next Section to calculate non--thermal emission from galaxy clusters.

We adopt a simplified situation: Figure ~\ref{fig:schema} shows the chain of 
physical processes that we consider and their interplay.
We assume that the thermal IGM is magnetized and, at time $=0$, consider (only) 
relativistic protons, with initial spectrum $N_p(p)=K_p p^{-2.6}$. 
The presence of relativistic
and thermal protons determines the initial efficiency of injection 
of secondary particles in the IGM (Sect.~3.4), whose initial spectrum 
is calculated assuming stationary conditions (e.g. Dolag \& Ensslin
2000; i.e. by taking $D_{pp}=0$ and $\partial N/\partial t =0$ in 
Eq.~\ref{elettroni}).

\noindent
Following Brunetti \& Lazarian (2007), 
we assume that compressible turbulence is injected at large scales 
and develops a quasi--stationary spectrum (Eq.~\ref{cascade_spectrum}) 
due to the interplay between non-linear wave-wave interaction and 
collisionless dampings at smaller scales (Sect.~3.1).
Following Brunetti \& Lazarian (2007) we consider a
time--independent damping, $\sum \Gamma_i \simeq \Gamma_{th}$, 
that is obtained by assuming the 
(initial) physical properties of the IGM.
This is motivated by the fact that in our model
TTD dampings with relativistic particles 
are sub-dominant with respect to those with thermal IGM (see Cassano \& Brunetti 2005 and
Brunetti \& Lazarian 2007) and by the fact that the thermal properties of the
IGM are not greatly modified by turbulence.

\begin{figure*}
\includegraphics[width=\columnwidth]{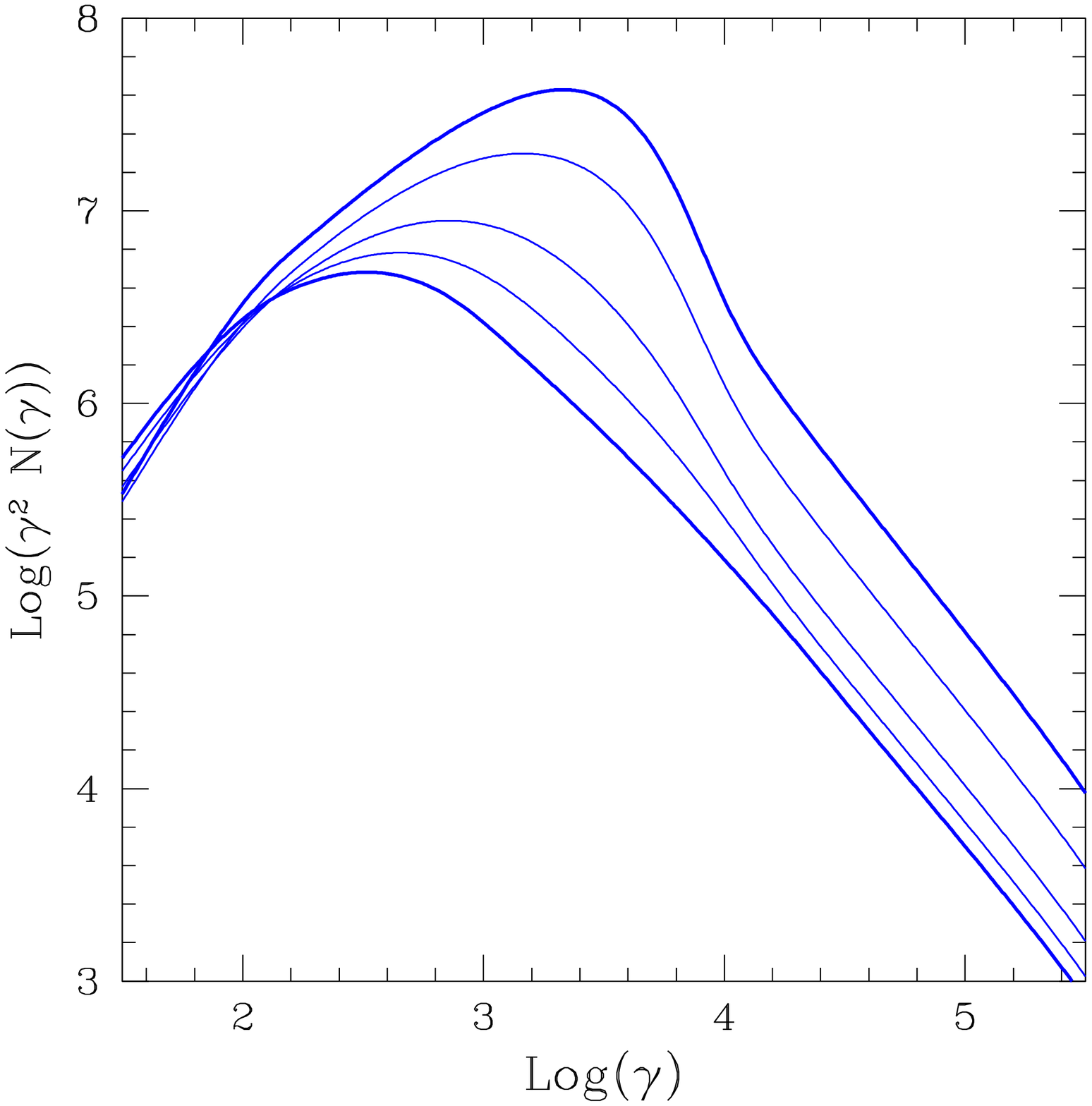}%
\includegraphics[width=\columnwidth]{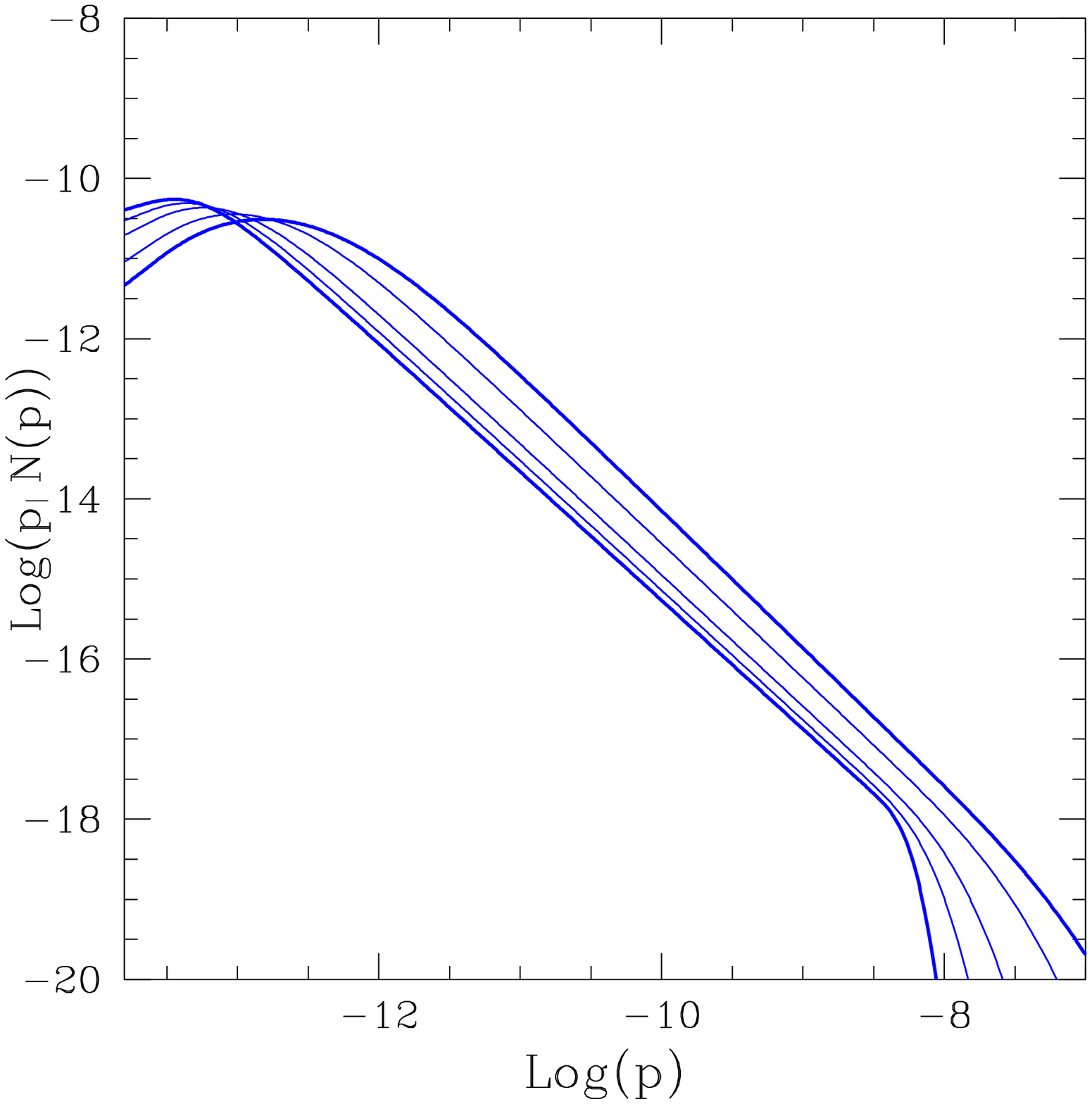}
\caption{The time evolution of the spectrum of electrons (left panel) and
protons (right panel).
From bottom to top, calculations are shown after 0.3, 1, 2, 4, 6.5
$\times 10^8$ yrs of reacceleration.
Calculations are obtained at $z=0$, assuming $(V_L/c_s)^2 = 0.22$, 
with injection scale $L_o = 300$ kpc, $T=10^8$ K, thermal
density $n_{th}=2.64 \times 10^{-3}$cm$^{-3}$ and $B = 4.69 \mu$G.}
\label{fig:evolution1}
\end{figure*}

\noindent
In our calculations we follow self-consistently the re-acceleration of 
relativistic protons due to TTD with the compressible turbulent-modes, 
the generation (and its evolution with time due to the evolution of the 
spectrum of protons)
of secondary electrons and positrons through collisions between 
these protons and the IGM, and the  
reacceleration of the secondaries (see Fig.~\ref{fig:schema}).

\noindent
Theoretically 
we expect that magnetic field can be amplified by turbulence in the IGM (e.g.
Subramanian et al 2006), although as a necessary simplification we do 
not include that amplification process in our calculations. 
The main reason is that the spectrum of the reaccelerated relativistic
electrons is expected to evolve more rapidly than the magnetic field in the
IGM (e.g. Cassano 2010). We also stress that present data do not show a clear connection
between the magnetic field properties and cluster dynamics (e.g.,Clarke et 
al 2001; Govoni 2006) that
leaves the process of magnetic field amplification still poorly constrained.

Turbulent acceleration can be thought as the combination of a systematic
effect, that causes the boosting of the spectrum of particles at higher
energies, and a stochastic effect, that causes a broadening of the
spectrum with no net acceleration (e.g. Melrose 1968; Petrosian 2001).
The time-scale of the systematic acceleration is :

\begin{equation}
\tau_{acc} = {{ p^3 }\over{\partial p^2 D_{pp} / \partial p }} =
{{ p^2 }\over{ 4 D_{pp} }}
\label{tauacc}
\end{equation}

\noindent
that does not depend on the particle energy in the case of TTD 
acceleration (Sect. 3.2), 
provided that the spectrum of 
compressible turbulence is isotropic (Sect. 3.1).

\noindent
By considering a reference value of the acceleration time due to TTD resonance
in the IGM, $\tau_{acc} \approx 10^8$ yrs (e.g., Cassano \& Brunetti 2005; 
Brunetti \& Lazarian 2007), Coulomb\footnote{In the
external regions of galaxy clusters Coulomb losses are less severe and
lower energy electrons can be reaccelerated (Fig.\ref{fig:times_ep})}
and radiative losses in the IGM prevent the reacceleration of
electrons with energies $E < 10$ MeV and $E > 10$ GeV, respectively 
(Fig.\ref{fig:times_ep}).

On the other hand, TTD resonance in the IGM may reaccelerate supra-thermal 
protons up to high energies, and consequently
in our model the energy density of relativistic protons increases with mergers, 
similarly to the energy of the thermal IGM.
By considering a toy scenario where these protons are simply injected in the
cluster volume, with initial energy density $\epsilon_{RC}(0)$, and then reaccelerated 
by MHD turbulence during major cluster mergers, the rate of increase of their
energy is :

\begin{equation}
d \epsilon_{CR}/dt \sim 
\int 
dk W(k,t) \Gamma_{CR}(k,t)
\label{protoni1}
\end{equation}

\noindent
where the TTD damping of compressible MHD turbulence by relativistic protons 
in case $\beta_{pl} >> 1$ (from Eq. 31 in Brunetti \& Lazarian 2007) is :

\begin{equation}
\Gamma_{CR}/\omega_r \simeq {{\pi^2}\over{4}} {{c_s}\over{c}} 
{{ \epsilon_{CR} }\over{ \epsilon_{th} }}
{{ \sin^2 \theta}\over{| \cos \theta |}} 
\left\{ {{\partial \hat{f} }\over{\partial p}} {{p}\over{\hat{f}}} 
\right\}
\label{dampingHbetacr}
\end{equation}

\noindent
where $\left\{ ...\right\} = s+2$ assuming a power law energy distribution of 
relativistic protons $N(p) \propto  p^{-s}$.
By assuming a "sonic" turbulent--forcing, $I_o \tau_{cl} \approx \epsilon_{th}$,
where $\tau_{cl} \sim 3-6$ Gyrs is the life--time of massive clusters,
and a typical merging history of massive clusters (eg., Cassano \& Brunetti 2005
and ref. therein),  
we expect $\epsilon_{CR} \leq$ few percent of the thermal energy density, 
provided that relativistic protons are injected in the IGM with an ``initial'' 
energy density $\epsilon_{CR}(0)/\epsilon_{th} \sim 0.001-0.01$\footnote{this is 
obtained by using the approximate scalings ${{ k W(k) }\over{ \tau_{kk} }} \approx I_o$
in Eq.\ref{protoni1}}.

In Figure \ref{fig:evolution1} we report the time evolution of relativistic 
electrons and protons in a hot, $T \sim 10^8$ K, IGM
by assuming $(V_L/c_s)^2 =$0.22, in which case the TTD acceleration 
time is $\tau_{acc} \sim 10^8$yrs.
Radiative losses prevent the acceleration of relativistic electrons 
above a maximum energy, $\gamma_{max} \approx 10^4$, producing a bump in the spectrum of 
electrons (Fig.~\ref{fig:evolution1}a). 
The number density of high energy electrons increases with time also for 
$\gamma > \gamma_{max}$ (Fig.~\ref{fig:evolution1}a),
that is because the injection rate of secondary electrons is enhanced 
with time as protons
are accelerated (Fig.~\ref{fig:evolution1}b).
Overall the mechanism is very efficient : on one hand the acceleration
of relativistic protons enhances the injection rate of secondary
electrons, at the same time an increasing number of secondary electrons 
accumulates at energies $\approx \gamma_{max}$ where cooling 
is balanced by acceleration.
The combination of these two effects boosts the spectrum of electrons 
at energies $\gamma \approx \gamma_{max}$.

\begin{figure}
\includegraphics[width=\columnwidth]{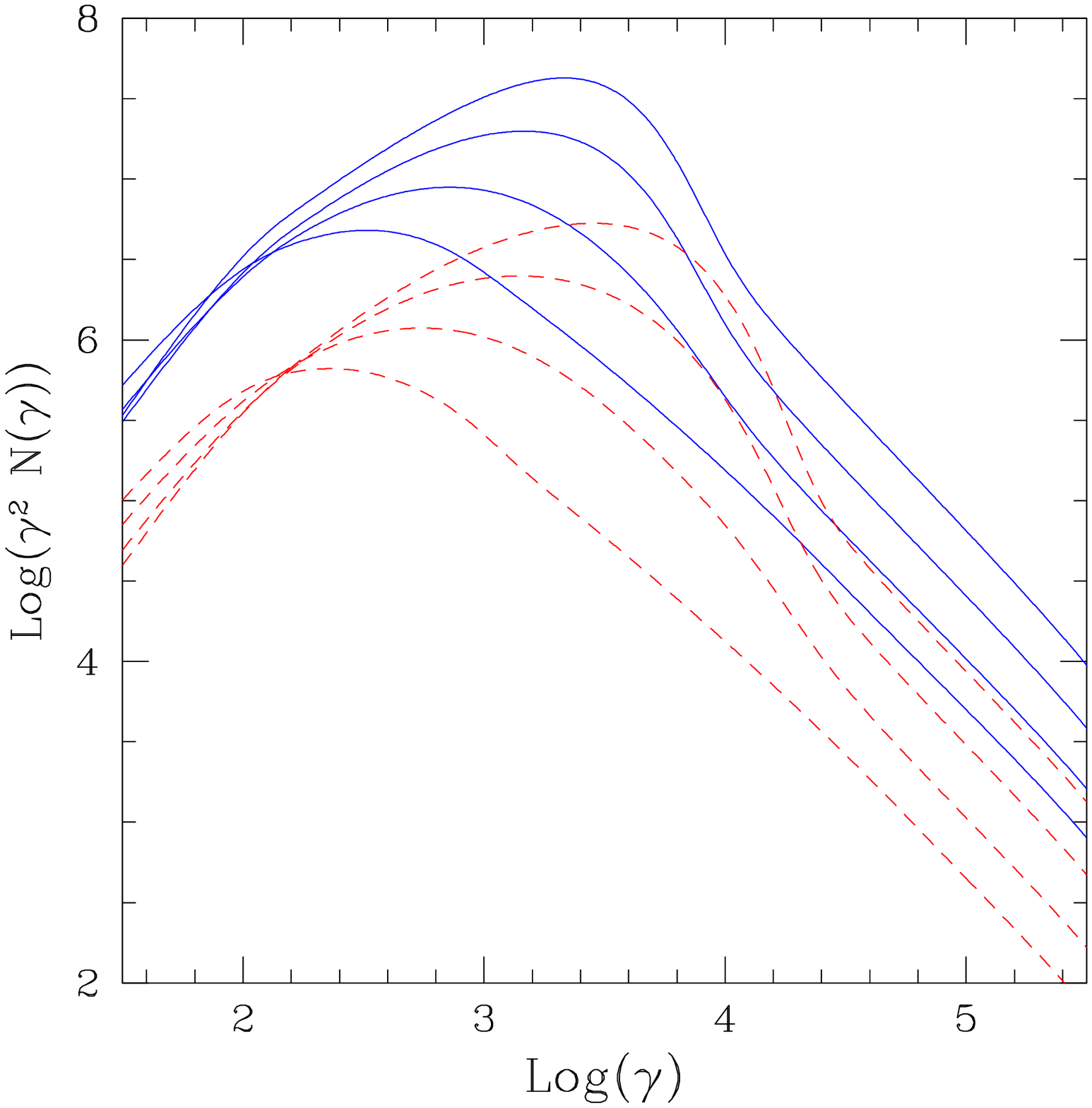}%
\caption{ 
The time evolution of the spectrum of electrons in two different environments:
in the clusters central regions (solid lines), assuming $n_{th}= 2.64 \times
10^{-3}$cm$^{-3}$
and $B= 4.69 \mu$G, and in the clusters external regions (red dashed lines),  
assuming $n_{th}= 0.56 \times 10^{-3}$cm$^{-3}$ and $B= 2.16 \mu$G.
From bottom to top, calculations are shown after 0.3, 2, 4 and 6.5 
$\times 10^8$ yrs of reacceleration.
Calculations are obtained at $z=0$, assuming $(V_L/c_s)^2 = 0.22$, with 
injection scale $L_o = 300$ kpc, $T=10^8$ K.}
\label{fig:evolutionsh1vs6}
\end{figure}

In Figure \ref{fig:evolutionsh1vs6} we report the evolution of the spectrum of 
electrons with time assuming physical conditions that 
span the volume occupied by radio halos, from cluster core  
(solid lines) to 1 Mpc distance from the cluster center (dashed lines); 
$(V_L/c_s)^2 =$0.22 is assumed in both cases.
Figure \ref{fig:evolutionsh1vs6} highlights the effect of decreasing 
radiative and Coulomb losses 
on the spectrum of the reaccelerated electrons : the 
boosting of the particle spectrum increases 
in the external (Mpc-distance) regions because Coulomb losses are less severe
(which allows the reacceleration of the bulk of secondary electrons) 
and because synchrotron losses become sub-dominant
with respect to inverse Compton losses due to the scattering of the CMB photons.

\subsection{The non-thermal spectrum of turbulent galaxy clusters}

In this Section we calculate the non thermal spectrum 
from galaxy clusters assuming 
reacceleration of primary protons and of their secondary products 
by compressible MHD turbulence in the IGM.

\begin{figure*}
\includegraphics[width=\columnwidth]{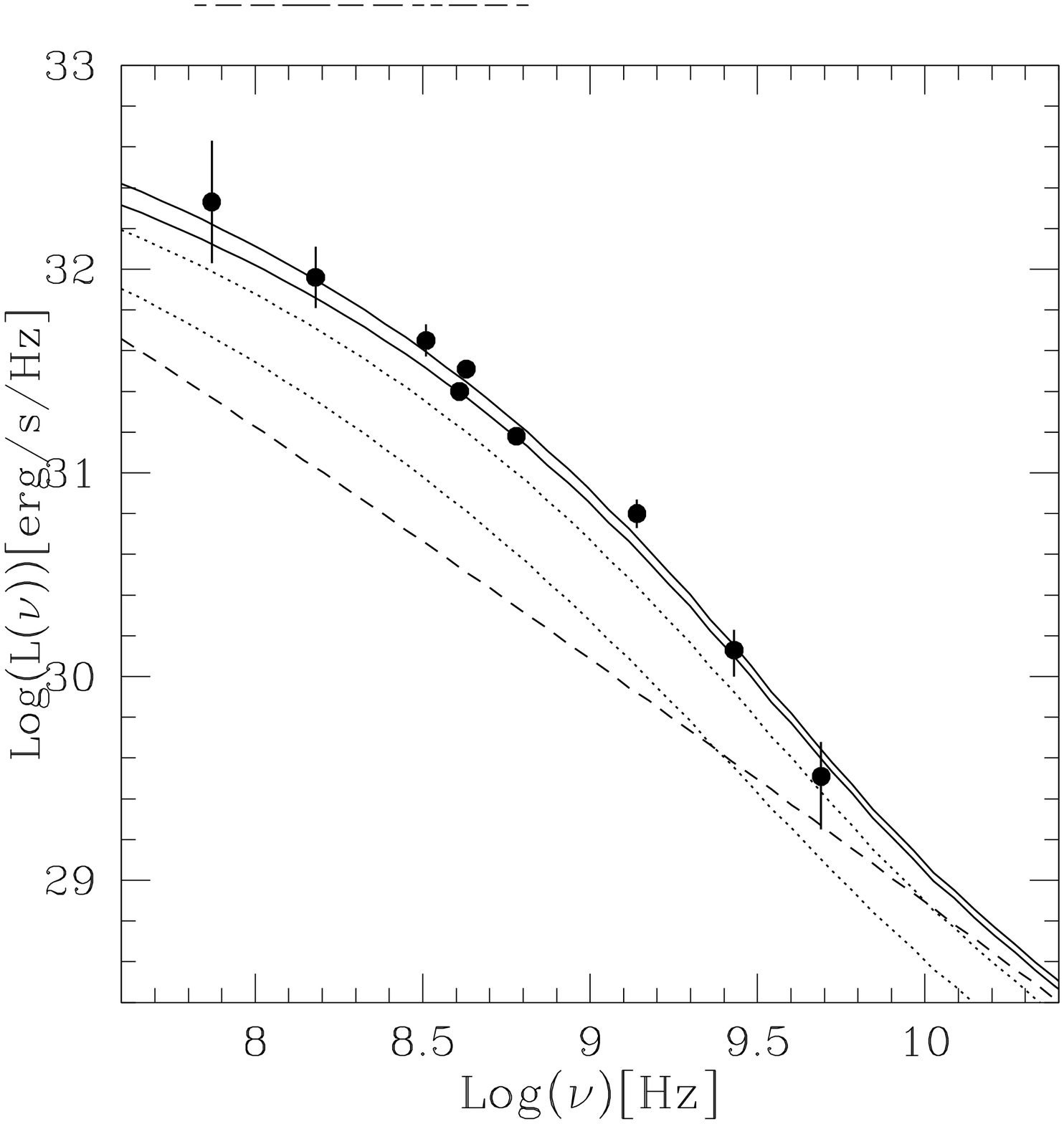}%
\hspace*{\columnsep}%
\includegraphics[width=\columnwidth]{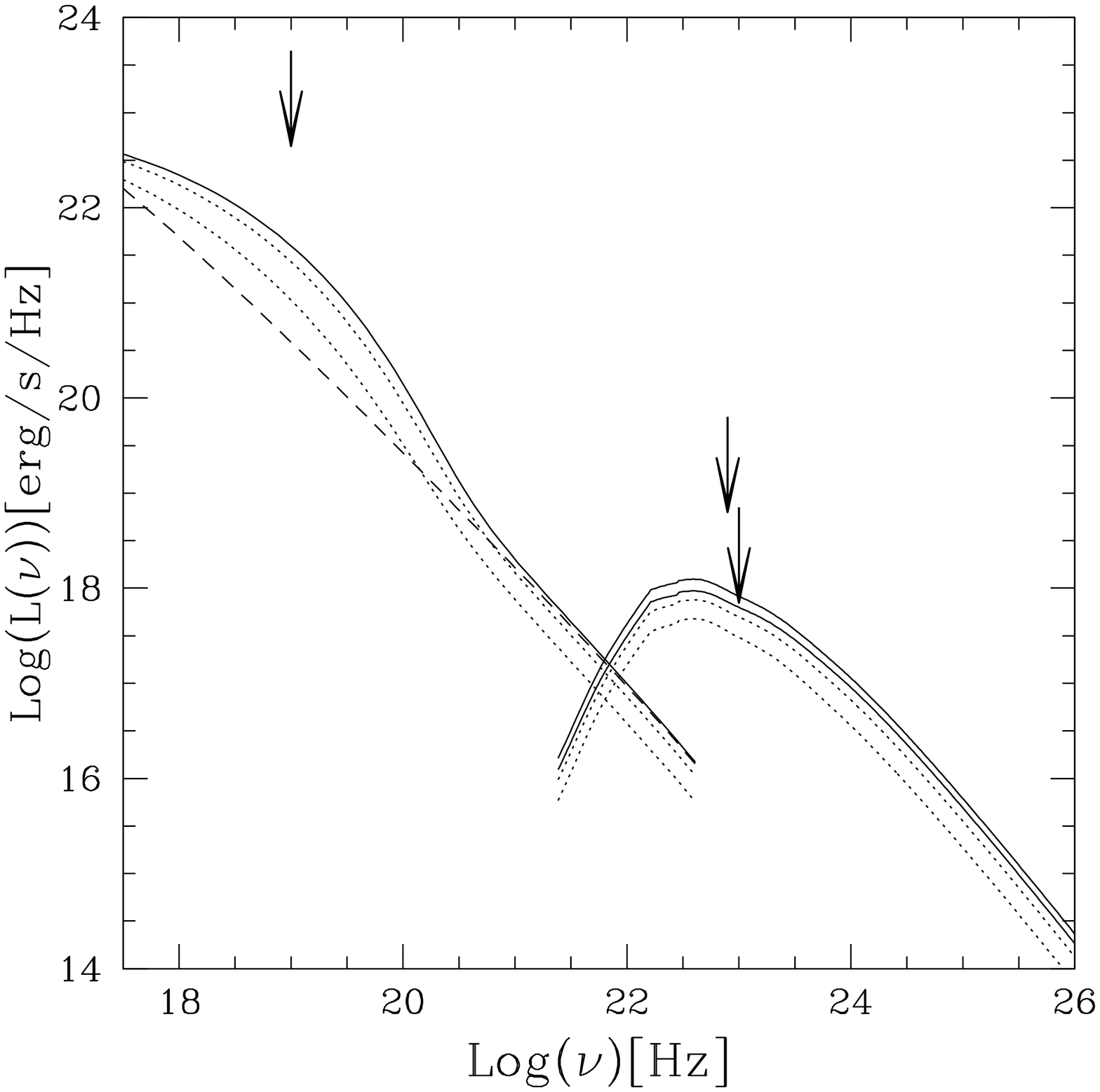}
\includegraphics[width=\columnwidth]{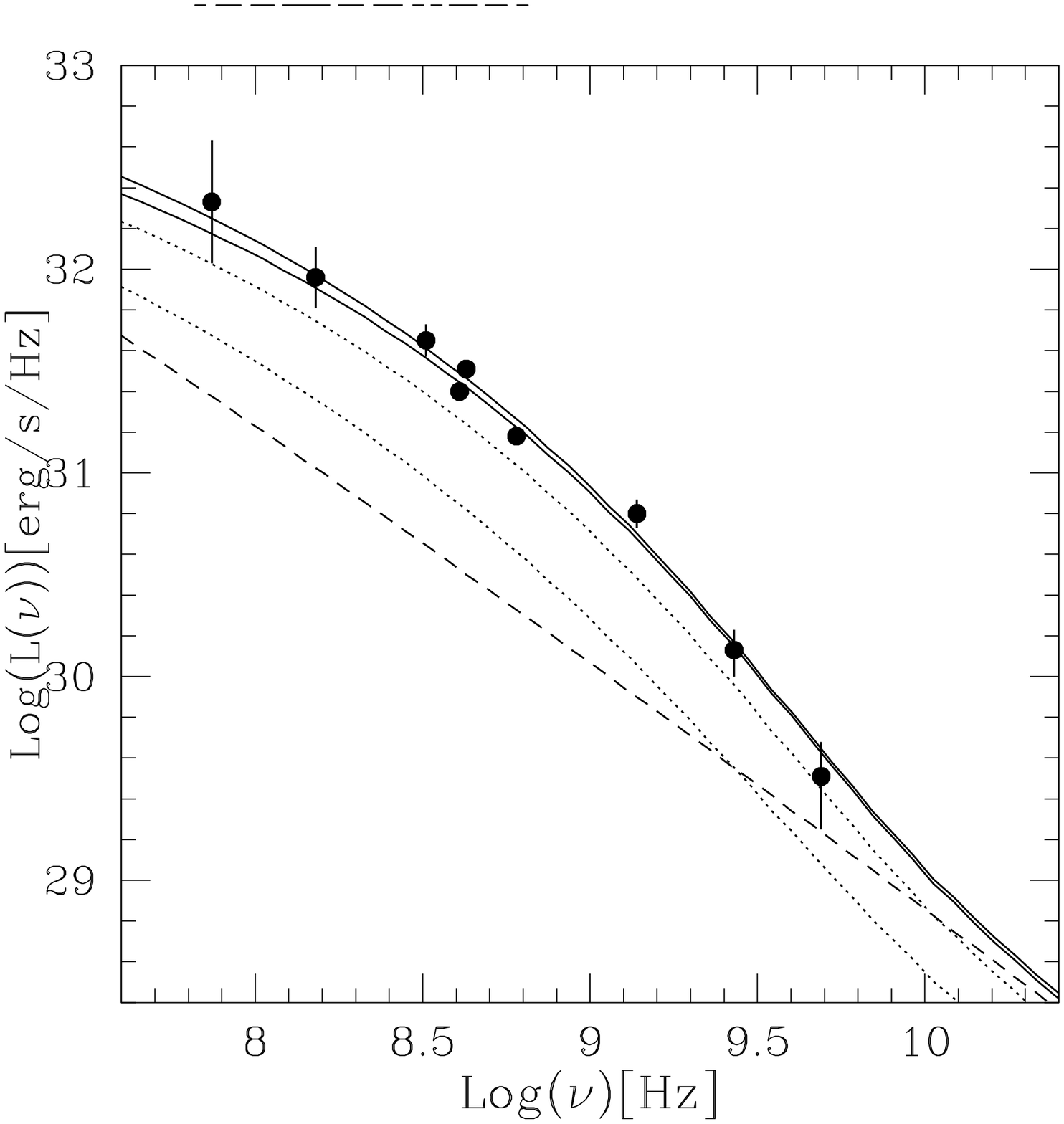}%
\hspace*{\columnsep}%
\includegraphics[width=\columnwidth]{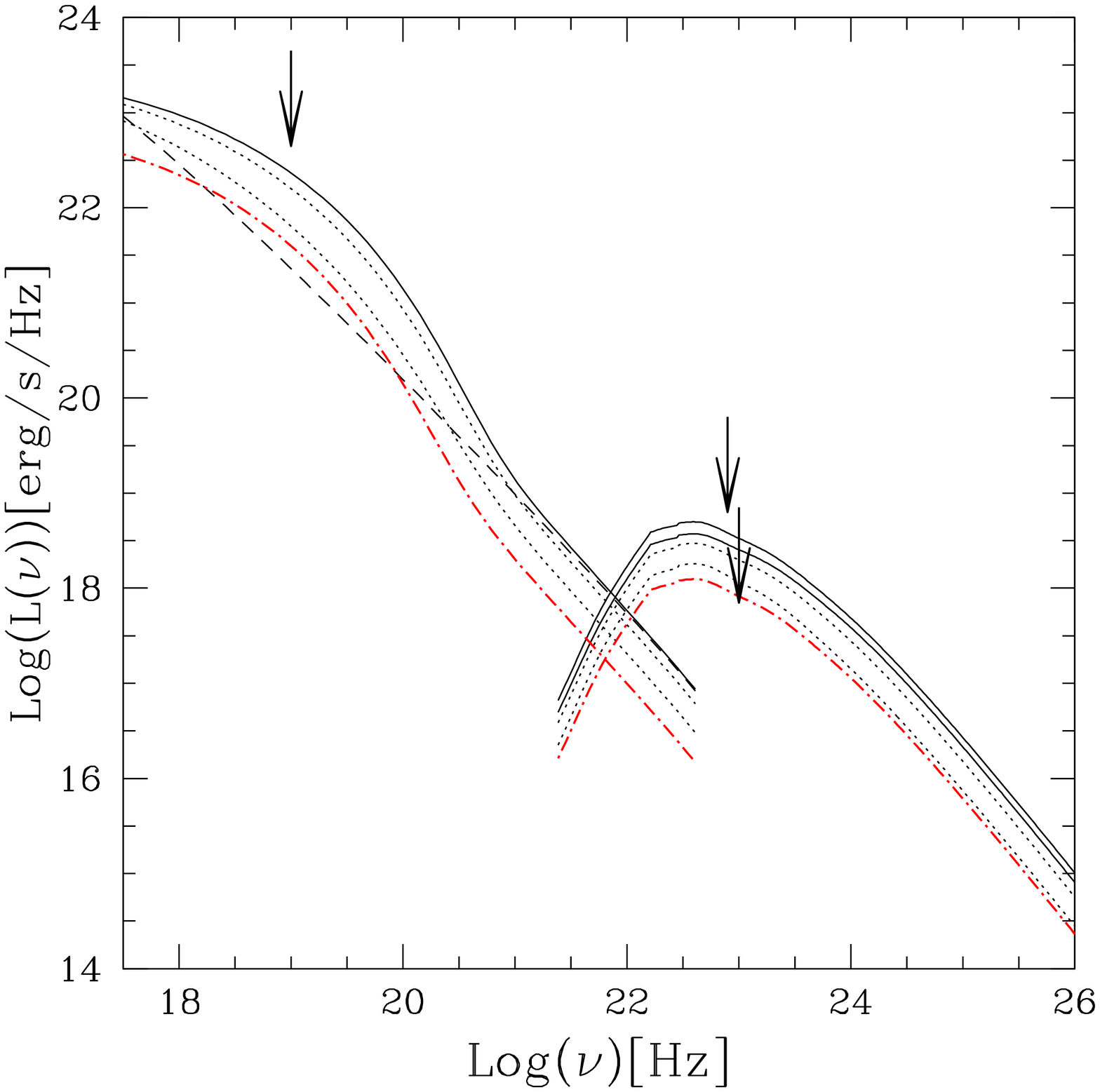}
\caption{The radio (left) and high energy (right) emitted spectra from 
the Coma cluster (note that the SZ decrement
at higher radio frequencies is not included, see e.g. Donnert et al. 2010b). 
{\bf Upper panels} show our reference model (see text), {\bf lower panels}
refer to the case of a cluster magnetic field 2.5 times smaller than that in
our reference model (for a better comparison 
red dot--dashed lines in the right panel show the IC and gamma-ray emission
expected in our reference model).
Solid lines show the emitted spectra obtained after 6.5 $\times 10^8$ yrs 
of reacceleration (upper and lower lines in the synchrotron and gamma-ray
spectra refer to the emission within 4.5 and 3 core radii, respectively, 
while X--rays are calculated within 4.5 core radii), dashed lines 
show the emitted spectra obtained
by assuming that MHD turbulence is dissipated in the cluster.
Dotted lines show the emitted spectra at intermediate stages 
of reacceleration : 
after 2 and 4$\times 10^8$ yrs, from bottom to top respectively.  
Radio data--points are taken from Thierbach et al (2003), the 
hard X-ray upper limit
is from Suzaku observations (Wik et al 2008) and the EGRET 
and FERMI upper--limits
are taken from Reimer et al (2003) and Ackermann et al (2010), respectively.
Calculations assume $(V_L/c_s)^2 \sim 0.18$ for our reference model
and 0.2 for the lower panels ($L_o = 300$ kpc is assumed in all calculations).
For reference, the energy of the population of relativistic protons "measured" 
after 0.65 Gyr of
reacceleration (solid lines) is 0.035 and 0.25 times that 
of the thermal cluster
in our reference model and lower panels, respectively.
}
\label{fig:brunetti07}
\end{figure*}

\subsubsection{A toy model for the Coma cluster}

The Coma cluster hosts the best studied, prototype, giant radio halo 
(Willson 1970; Giovannini et al 1993). 
Constraints on the high energy emission from the Coma cluster are presently 
available from
hard X-ray (BeppoSAX, Fusco-Femiano et al 1999, 2004, Rossetti \& Molendi 2004; 
RXTE, Rephaeli et al 1999; INTEGRAL, Eckert et al 2007, Lutovinov et al 2008; 
Suzaku, Wik et al 2009; Swift-BAT, Ajello et al 2009) and 
gamma ray (EGRET, Reimer et al 2003;
FERMI, Ackermann et al 2010; HESS, Aharonian et al 2009b;
VERITAS, Perkins 2008) observations.
It is thus a natural first step to compare our model expectations 
with the spectral energy distribution (SED) of the non thermal emission from
the Coma cluster.

We assume a simplified model for the
thermal gas distribution in the Coma cluster that is anchored to 
the observed beta--model profile (e.g. Briel et al 1992, 
considering $\Lambda$CDM cosmology).
The gas temperature, $k_B T \simeq 8$ keV (David et al 1993), 
is assumed to be constant on Mpc-scale.

\noindent
The magnetic field in the Coma cluster, and its spatial distribution, 
is a crucial ingredient in our modeling of the synchrotron
(radio halo) properties. We adopt the recent results by Bonafede et al (2010) 
that carried out a detailed analysis of the Rotation Measures of a sample 
of extended cluster radio galaxies in the Coma cluster.
The best fit to their data implies a magnetic field
$B(r) \propto \sqrt{n_{th}}$ with a central value $B(0) \simeq 5 \mu$G.

Once the thermal gas and magnetic field properties are anchored to present data, 
the free parameters in our calculations are (i) the energy density 
(and spatial distribution) of relativistic protons (at time $=0$)\footnote{we assume
an initial spectrum of protons $N_p(p)=K_p p^{-2.6}$}, and (ii) 
the injection rate (and scale) of compressible turbulence in the IGM, $I_o$.

\noindent
In our reference toy model we assume (i) a flat spatial distribution of relativistic 
protons on the halo--scale, $r \sim 3\,r_c$, and a constant ratio between relativistic
and thermal particles energy densities at larger distances, 
and (ii) a specific injection rate of turbulence, $I_o/\rho$, constant on the 
halo--scale. 
Assumption (i) is motivated by the fact that, similarly to other giant halos, 
the Coma radio halo has a very broad synchrotron--brightness distribution 
(Govoni et al 2001; see also discussion in Cassano et al 2007 and Donnert et al
2010a) implying a very broad spatial distribution of relativistic protons 
on the halo--scale. Assumption (ii) is motivated by hydrodynamical and MHD 
cosmological simulations that found very extended turbulent regions in simulated 
galaxy clusters (Sunyaev et al 2003; Dolag et al. 2005; Vazza et al 2009a; 
Paul et al 2010). 

\begin{figure}
\includegraphics[width=\columnwidth]{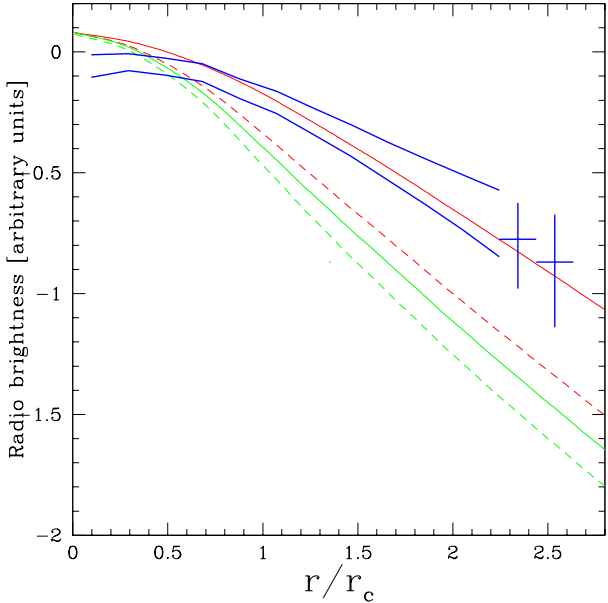}%
\caption{Radio brightness profile of the Coma radio halo (in arbitrary units).
The observed profile (tick--blue lines and crosses) is taken from Govoni et 
al (2001) observations.
Solid lines show the brightness profile at 330 MHz predicted by our reference 
model (red line) and by the model with lower magnetic field (green line).
Dashed lines refer to the brightness profiles of both models (same color code)
considering the case where turbulence is dissipated.}
\label{fig:brightness}
\end{figure}

In Figure \ref{fig:brunetti07} (upper panels) we show the expected SED emitted
by the Coma cluster (from $r \leq 3-4.5\, r_c$) 
assuming our reference model (see caption).
We find that the radio spectrum of the Coma radio halo can be well reproduced by 
assuming a total energy content of
relativistic protons, on the halo--scale, $E_{CR} \sim 3.5$\% of the thermal 
gas (see caption for details).
The presence of a break in the spectrum of the Coma radio halo at higher frequencies 
has been interpreted as a signature
of turbulent acceleration\footnote{more recently Donnert et al. (2010b) have 
shown that the observed steepening cannot be due 
to the SZ decrement due to the hot gas in the
central Mpc region of the cluster}, since it implies  
a corresponding break in the spectrum 
of the emitting electrons at the energy where turbulent acceleration is balanced 
by radiative losses
(e.g. Schlickeiser et al 1987; Brunetti et al 2001; Petrosian 2001).
Figure \ref{fig:brunetti07} demonstrates that also models considering the 
reacceleration of secondary particles can
explain the steepening of the Coma radio halo, although a tail at higher frequencies 
due to freshly--injected
secondary electrons shows up in the (emitted) spectrum\footnote{a similar conclusion
comes from calculations of reacceleration of secondary particles 
in the Alfvenic case (Brunetti \& Blasi 2005)}. 
For a consistency check, in Figure \ref{fig:brightness} we also show that, although 
very simplified, our reference model 
allows us to (roughly) reproduce the observed radio brightness profile of the Coma 
halo (although the predicted profile is still slightly steeper than the observed one).
Such expected--broad synchrotron profile is due to the combination of the flat spatial
distribution of relativistic protons with the increasing efficiency of reacceleration 
of secondary electrons at larger distances from the cluster centre (see 
Fig. \ref{fig:evolutionsh1vs6}).

The gamma ray emission from the cluster (from $r \leq 3-4.5\, r_c$ , see caption) 
is dominated by the decay process of  $\pi^0$ (Fig.~\ref{fig:brunetti07}) and
is expected at about 10 percent level of the present gamma--ray upper limits 
(FERMI, Ackermann et al 2010).
In our model the energy content of relativistic protons that is necessary to 
reproduce the observed luminosity and brightness profile 
of the Coma radio halo is much smaller than that from calculations based on
pure hadronic models that assume a similar magnetic field in the Coma cluster
(Donnert et al 2010b), implying also a much smaller gamma--ray luminosity.

\noindent
For seek of completeness, in Figure \ref{fig:brunetti07} (lower panels) we show the 
expected SED of the Coma cluster by assuming a magnetic field 
in the cluster 2.5 times smaller than that inferred from Bonafede et al.(2010) 
(see caption).
In this case the properties of the Coma radio halo can be reproduced by assuming an 
energy content of relativistic protons
$E_{CR} \sim 25$\% of the thermal energy, i.e. 7-8 times larger than that of our 
reference model, and the expected gamma ray emission increases significantly. 
Interestingly, in this case there would be a chance to detect the Coma cluster in 
the next years with the FERMI telescope.

\noindent
Remarkably, on the other way round 
Figure \ref{fig:brunetti07} (lower panels) demonstrates the 
importance of present gamma ray
upper limits: in the turbulent reacceleration picture,
the magnetic field in the central regions of the Coma cluster cannot be 
significantly smaller than about 2 $\mu$G, provided that hadronic collisions are the 
main source of the (seed) electrons in the cluster volume.

In the case of our reference model, the inverse Compton emission in the hard 
X-ray band is expected
at a $\sim$percent level than the present Suzaku upper limit and a 
detection of the Coma cluster 
will be challenging even with the future hard X-ray experiments (e.g. ASTRO-H, Nustar).
On the other hand, by assuming a magnetic field in the cluster 2.5 times smaller than 
that inferred from 
Bonafede et al.(2010) the expected inverse Compton emission increases significantly 
and there would be a chance to detect 
the Coma cluster with future hard X-ray telescopes.
However, by assuming a magnetic field substantially smaller than 
that of our reference model, we find that the predicted 
radio profile is much steeper than the observed one (Fig.~\ref{fig:brightness}).

\subsubsection{Transient and long-living spectral components in galaxy
clusters}

In the context of our model the non-thermal emission from galaxy clusters
is a mixture of two main spectral components :
a {\it long-living} one that is emitted by the chain of secondary particles 
continuously generated by collisions between thermal and long-living (several Gyrs)
relativistic protons,  
and a {\it transient} amplification of the SED that appears when relativistic
particles are reaccelerated by the MHD turbulence generated (and then
dissipated) in connection with cluster mergers (see also Brunetti et al 2009a); 
in this scenario the idea is that the 
{\it transient} synchrotron component generates 
the observed radio halos.

\noindent
The non-thermal spectrum of dynamically relaxed galaxy clusters should be 
mainly due to the long-living spectral component.
An example of this spectral component is shown in Figure \ref{fig:brunetti07} 
(dashed lines) where we report the synchrotron, 
inverse Compton and $\pi^o$ emission calculated by assuming that
MHD turbulence in the Coma cluster is dissipated.

\noindent
The main effect of turbulent-reacceleration is to produce a 
amplification of the synchrotron (radio), inverse Compton (hard X) and 
$\pi^o$ (gamma) emission in merging clusters (Fig. \ref{fig:brunetti07}), 
while the effect of
the dissipation of MHD turbulence in those clusters that become more relaxed is
to suppress (only) the radio and the hard X-ray emission\footnote{Also the radial
synchrotron profile in non--turbulent clusters is steeper than that in turbulent
clusters (Fig. \ref{fig:brightness})}. 
Consequently in our model we expect a tight correlation between
clusters dynamics and the radio (radio halos) 
and hard X-ray properties of galaxy clusters,
but do not expect a tight correlation between gamma rays and cluster dynamics.

An important issue is the evolution of the radio halos in connection with
cluster mergers and the difference between the radio properties of merging
and relaxed clusters.

\noindent
In the "classical" scenario, where turbulence reaccelerates 
primary electrons, the boosting of the synchrotron and hard X-ray emission
during cluster mergers can be extremely large (eg. Fig.~5 in Brunetti et al 2009b; 
Fig.~6 in Cassano 2010).
Contrary to that, in our model 
the presence of relativistic protons in the IGM and the chain of secondary-particles 
decay induced by pp collisions put stringent constraints 
to the radio--evolution of galaxy clusters, since long--living spectral components are
generated for about a Hubble time.
Our calculations exploit this point and unveil the interplay between transient 
and long-living spectral components.
In our model the boosting of the radio and hard X-ray luminosities due 
to reacceleration is fairly
constrained, since it is essentially
due to the reacceleration of secondary electrons from the energies where the injection
spectrum of secondary electrons peaks (or from the energy where 
reacceleration is stronger than Coulomb
losses) to the energies necessary to electrons to radiate synchrotron emission in the
radio band (provided reacceleration is sufficient to accelerate particles up to
these energies). 
This is $\delta P / P \leq (E Q^{\pm}(E))_{E_l} / (E Q^{\pm}(E))_{E_h}$, where
$E_l$ and $E_h$ are about 100 MeV and 1 GeV respectively, that implies
$\delta P /P \sim 10-15$ for typical values of the slope of 
the proton spectrum, $s = 2.4 - 2.8$.

\noindent
The predicted level of 
amplification and suppression of radio emission in galaxy clusters and
its connection with cluster dynamics can be constrained by radio and X-ray observations.
The recent radio follow up of a complete X--ray sample of
galaxy clusters, the "GMRT Radio Halo Survey" (Venturi et
al. 2007, 2008), leads to the discovery of a {\it bi-modal}
behaviour of the clusters radio properties (Brunetti et al 2007, 2009b): 
radio halos are found in the most disturbed systems and trace 
the $P_{1.4}-L_X$ correlation, on the other hand the majority of clusters 
are less disturbed and "radio quiet", with the limits to their radio 
luminosities about 10 times smaller than the radio luminosities of halos.
As a matter of fact, 
our reacceleration 
model predicts that the level of the long-living synchrotron radiation 
in "radio quiet" clusters is similar to that of the upper limits from 
current GMRT radio observations, on the other hand the "classical" reacceleration
scenario, where secondaries do not play a role, would predict a larger gap
between radio halos and "radio quiet" clusters.
Consequently future (deeper) radio observations may test the role of relativistic 
protons and secondary particles.

A second important issue is the possibility to detect gamma ray emission from
galaxy clusters.
In our reacceleration model gamma ray emission is a spectral component 
common in galaxy clusters, it essentially depends on the integrated merging history of 
clusters rather than to the dynamical status of a given cluster at the
epoch of the observation; this is similar to classical hadronic 
models (eg., Blasi et al 2007 for review) and indeed our model
evolves into these models when MHD turbulence in the IGM is dissipated.

\noindent
In our "reference" model--configuration we expect that 
gamma ray emission from galaxy clusters is difficult to 
detect in next years with present facilities.
However, assuming a smaller magnetic field in the IGM, a larger number of relativistic
protons must be assumed to explain radio halos and there is a chance to detect
gamma rays from clusters in next years.
As in the case of hadronic models, 
we expect that gamma ray emission scales with thermal properties as :

\begin{equation}
L_{\gamma} \propto \rho_{IGM}^2 M T 
\langle {{\epsilon_{CR}}\over{\epsilon_{th}}} \rangle
\label{lgamma}
\end{equation}

\noindent
where $\langle ... \rangle$ indicates emission weighted quantities in the cluster
volume. In the most simple situation where we assume that 
$\langle {{\epsilon_{CR}}\over{\epsilon_{th}}} \rangle$ does not depend on cluster
mass, Eq.\ref{lgamma} implies a quasi linear scaling between gamma 
and (thermal) X-ray luminosities 
of galaxy clusters, $L_{\gamma} \propto L_X^{1.3}$\footnote{Here we have considered 
$L_{X} \propto \rho_{IGM}^2 M \sqrt{T}$,
and assumed the empirical scaling $L_{X} \propto T^{2.98}$ (eg. Reiprich 2001).}
(see also Pfrommer 2008 and Donnert et al 2010a for more detailed calculations based
on cosmological simulations).

\section{Discussion and Conclusions}

If MHD turbulence is generated in galaxy clusters during massive merger events,
relativistic protons can be reaccelerated enhancing the rate of generation 
of secondary particles in the IGM.
Consequently, for a self-consistent treatment of the process, protons and 
their secondary products must be taken into account
in the calculations of particle reacceleration by MHD turbulence in galaxy clusters.

\noindent
In this paper we have assumed the picture of MHD turbulence in galaxy clusters
as described in Brunetti \& Lazarian (2007) to calculate the acceleration of primary
protons and secondary particles by compressible MHD turbulence.

Upper limits to the gamma ray emission from galaxy clusters and radio upper limits
to the cluster-scale synchroton emission in clusters without radio halos
constrain the
energy density of relativistic protons in the central Mpc regions of galaxy clusters
to less than a few percent of that of the IGM (Reimer et al 2004; 
Brunetti et al 2007, 2008; Aharonian et al 2009a,b; Ackermann et al 2010).

We have shown that a population of relativistic protons consistent with 
the above limits may be sufficient to generate the observed radio halos 
in merging clusters (including their observed brightness profiles and spectra) 
via the reacceleration of their secondary electrons
by compressible MHD turbulence, 
provided that the spatial distribution of the relativistic protons is relatively flat 
and that the magnetic field in galaxy clusters is at the level 
derived from Rotation Measurements.

\noindent
The main consequence of this theoretical scenario is that the non--thermal SED of
galaxy clusters is given by the interplay of a transient component, generated by
the reacceleration of secondary particles by MHD turbulence during cluster mergers
(that generates radio halos),
and a long--living component, that is generated by the secondary particles that are
continuously injected by pp collisions in the IGM.
In this case we expect a tight connection between radio halos and cluster mergers,
as well as an amplification of the level of hard X-ray emission in merging clusters,
while we expect no tight correlation between gamma rays and cluster dynamics.

\noindent
At the same time, we also expect that diffuse radio emission, due to secondary particles, 
must be common in galaxy clusters at a level that is about one order of magnitude below 
that of nowadays observed radio halos, i.e. at the same level of the upper limits 
derived by present radio observations of "radio quiet" clusters.
This expectation can be tested by future (deeper) radio observations of "radio quiet"
clusters.

\noindent
The level of gamma ray emission from nearby, massive, galaxy clusters is expected 
at about 10\% of the level of present upper limits, assuming a magnetic field strength 
in these clusters in line with present studies of RM. This implies that only future
telescopes (e.g. CTA) may lead to the detection of galaxy clusters in the gamma ray band. 
On the other hand, if the magnetic field in galaxy clusters is smaller, 
detection of galaxy clusters with FERMI could be possible in next years.
In this respect, in the context of our model, present FERMI upper limits for the Coma
cluster already provide a limit to the central value of the magnetic field 
$B(0) \geq 1-2 \mu$G.

\subsection{Model simplifications and future steps}

In our paper we focus on the role played by compressible 
MHD turbulence, essentially the fast modes. 
Our educated guess, motivated in Sect.2 and in Brunetti \& Lazarian (2007), is that 
these modes are the most relevant 
for the reacceleration of relativistic particles in galaxy clusters. 
These modes were also identified as major scattering agent for Galactic cosmic 
rays (Yan \& Lazarian 2002, 2004). 

\noindent
The Alfven modes and slow modes are not efficient 
for scattering if the turbulent energy is being injected at large scales 
(Chandran 2000, Yan \& Lazarian 2002). Their inefficiency stems from the both the 
spectra being steep in terms of parallel perturbations as well as fluctuations being 
very anisotropic (Goldreich \& Sridhar 1995, Lazarian \& Vishniac 1999, 
Cho \& Vishniac 2000, Maron \& Goldreich 2001, Cho, Lazarian \& Vishniac 2002). 
If Alfven modes are injected by instabilities they may have radically different 
properties from the modes of the large scale cascade. For instance, Alfven modes 
arising from particle streaming have slab structure. Such waves efficiently 
interact with energetic particles. Other instabilities, e.g. gyroresonance one, 
can produce slab waves (see Gary 1993). The instabilities producing slab Alfven modes 
may be induced by large scale compressible turbulence (see Lazarian \& Beresnyak 2006). 
Indeed it is worth mentioning that the mode composition at smaller scales, $l<<l_A$, 
in the IGM could becomes rather complex (e.g. Kato 1968; Eilek \& Henriksen 1984).

To what extend our calculations are accurate depends on our understanding of properties 
of IGM turbulence.  
We assumed that the damping of fast modes with thermal particles 
can be obtained considering a collisionless plasma. 
At the same time, it can be argued (see e.g., Lazarian et al. 2010) that the degree of 
collisionality of astrophysical plasmas can be underestimated if only Coloumb 
collisions are taken into account, as particles in plasmas can interact 
through the mediation of the perturbed magnetic fields.
The inevitable conclusion is that the collisionless formulae describing damping of 
these modes should be only applied to scales less than the mean free path, which is 
much shorter than the Coulomb mean free path. As a result, fast modes should be 
substantially less damped and be present on the scales which are much shorter 
than the earlier estimates, including those in this paper. 
The consequence of this is that a more appreciable portion of energy  
gets available for the acceleration of cosmic rays. 
Therefore, our present calculations may underestimate the 
efficiency of cosmic ray acceleration by turbulence. 
We plan to address the self-consistent problem elsewhere.     

Following Brunetti \& Lazarian (2007) in our paper we adopt 
quasi--linear--theory (henceforth QLT) to calculate particle 
acceleration by fast modes. 
In Brunetti \& Lazarian (2007) we indeed have shown that for fast modes in the
IGM it is $\langle \omega \rangle >> \langle \Gamma \rangle$, where 
$\langle  ... \rangle$ indicates angle--averaged quantities, that provides 
some justification to the use of QLT.
More recent studies in Yan, Lazarian \& Petrosian (2008) presented an
approach to particle acceleration that generalizes 
the QLT and allows to take into account the effect of large scale variations 
of magnetic field. 
As numerical simulations which use the data from the actual 
MHD turbulence simulations (Beresnyak, Yan, Lazarian 2010) support the theory, 
we believe that, in future, going beyond the standard QLT approach may provide a more
accurate description of the particle acceleration process, although we do not
expect a large change (see Yan et al 2008).

Finally a unavoidable simplification in our semi-analytical calculations
is that turbulence is homogeneous, in space and time, on the radio halo
volume, during cluster mergers. On the other hand, cosmological numerical simulations 
of galaxy clusters show a more complex situation where intermittent and patchy 
large--scale turbulent motions are generated during multiple
collisions between galaxy 
clusters (e.g. Vazza et al 2009a; Paul et al 2010). 
We believe that to be more realistic one may need to vary the intensity of 
turbulence driving to describe different parts of galaxy clusters.
Our idealized calculations of particle acceleration and evolution of compressible 
turbulence in the IGM provides a first step, implementing our formalism in detailed 
cluster--simulations is necessary to obtain a more reliable description of the 
morphology and spectral distribution of the non--thermal emission and of the connection 
between radio halos and cluster mergers.

\subsection{A comparison with the Alfvenic approach}

Earlier calculations of turbulent acceleration of primary protons and
secondary electrons in galaxy clusters focus on the reacceleration
by Alfv\'en modes (Brunetti \& Blasi 2005).
These studies first provided a description of the expected transient and long--living
spectral components in galaxy clusters (e.g., Brunetti et al 2009a).

\noindent 
It is well known that the damping of these modes is mainly due to 
the interaction (gyro--resonance) with relativistic particles, that provides 
the main motivation to explore this possibility.
On the other hand, as already mentioned, a complication of this approach is the anisotropy 
of Alfv\'en modes that develops when turbulence cascades from larger 
to smaller scales.
Consequently in Alfv\'enic models the injection of Alfv\'en modes 
"directly" at small scales, i.e. comparable with the gyroradius of high energy particles,
must be postulated, in which case it is also difficult to derive a overall 
picture connecting clusters mergers and the generation of these modes 
at such small scales. 

\noindent
A second issue in the modeling of Alfv\'enic acceleration of relativistic protons 
and of their secondary products is that the primary and secondary particles interact
with Alfv\'en modes with different scales.
This is because secondary electrons of energy $E_e$ are mostly generated by collisions 
between higher energy protons, $E_p \sim$ 10--50 $\times E_e$, and thermal targets
that implies that these 
secondaries interact with modes on scales 10--50 times smaller than 
those of the parent primary protons (by consider the gyroresonant conditions 
$k \sim e B / (c \, p \, \cos\theta)$).
Consequently the ratio between the transient and
long--living components in these models depends also on the spectrum of Alfv\'en modes.
This implies a larger degree of freedom in Alfv\'enic models with respect to the
more straightforward case treated in our paper where compressible MHD turbulence interact
with relativistic particles.

\subsection{Sources of primary electrons in a turbulent IGM}

In general, if relativistic protons and electrons are present in the IGM, 
we expect that the MHD turbulence, generated during cluster mergers, would
reaccelerate both these particles.
If we assume an efficient confinement of cosmic rays in galaxy clusters,
the unavoidable consequence of this scenario is that also the energy density 
of secondary products, due to pp collisions, should increase and  
consequently their contribution to the non-thermal clusters spectrum.

\noindent
In our paper we have addressed this problem under the most extreme (and
simplified) condition where only protons and their secondaries are 
present in the IGM; we note that under these conditions the ratio of the 
energy densities of relativistic protons and of the emitting electrons in 
the IGM is maximized. 
In this context, we have shown that, assuming turbulent reacceleration
at the level necessary to explain radio halos, the radio emission 
generated by secondary particles when turbulence is dissipated is 
consistent with present upper limits to the diffuse radio luminosity of 
clusters without radio halos.
Also we have shown that the gamma--ray emission from the decay of
secondary neultral--pions, that should be common in galaxy clusters and 
not tightly connected with their dynamical status, is expected at a level 
much smaller than that constrained from present upper limits with FERMI. 

\noindent
At the same time, however, under these conditions our calculations suggest 
that the difference between the cluster-scale radio emission of 
"turbulent" (merging) and "non--turbulent" (relaxed) clusters cannot be 
larger than about a factor ten, and that we expect to
detect diffuse radio emission potentially in "all" massive clusters as soon as much
deeper observations of clusters that are 
presently defined "radio--quiet" will become available.

On the other hand primary relativistic electrons should be present, at some
level, in the IGM.
It is well known that 
active radio galaxies may fill large volumes in the IGM with relativistic plasma, 
relativistic electrons age rapidly but they can be accumulated at hundred MeV energies 
for longer times, especially in the external regions (Fig.1).
Other sources of relativistic primary electrons that are usually considered in the
literature are Galactic Winds and Starburst galaxies (e.g., V\"olk \& Atoyan 1999) 
and shock waves (e.g., Ensslin et al 1998; Sarazin 1999;
Ryu et al 2003; Pfrommer et al. 2006; Skillman et al 2008;
Vazza et al 2009b), evidence for the
latter process come from the observations of radio relics that indeed suggest a 
connection between shocks and electron acceleration (or reacceleration)
in the IGM (e.g., Markevitch et al 2005 and Giacintucci et al 2008 for 
observations of a shocks--relics connection in clusters).

\noindent
In addition to these processes, another possibility that requires more attention 
is that the primary electrons can be "created" in the IGM. 
Magnetic reconnection presents a natural way for doing this, as electrons 
bounces back and forth between converging magnetic fluxes
can gain energy through the first order Fermi acceleration (Gouveia dal Pino 
\& Lazarian 2003).
Although it is generally believed that reconnection is a slow process, potentially
turbulence can significantly enhance the reconnection rate 
(Lazarian \& Vishniak 1999). Exploring this mechanism and its interplay with
particle reacceleration by MHD turbulence in the IGM may open new perspective
in our understanding of non--thermal cluster--scale emission and of its connection
with cluster mergers. We aim to discuss in detail this point in a future
paper, yet for completeness here we dedicate an Appendix to turbulent reconnection.

The presence of primary electrons in the IGM may significantly 
affect the SED of galaxy clusters and its evolution.
In particular if the number density of primary electrons 
is comparable to (or larger than) that of secondary particles, a smaller number
of relativistic protons is required by the our model to match the observed 
spectrum of radio halos. 
This has an impact on the expected gamma ray emission from galaxy clusters
that would be consequently smaller than that expected from the 
calculations presented in Sect.4.3.
A second important point is that if a population of 
reaccelerated primary electrons significantly
contributes to the observed radio halo emission, the expected 
ratio between transient and 
long living components in the radio and hard X-ray bands increases with respect to
that derived in Sect.4.3.
As a matter of fact, assuming that primary electrons are dominant with
respect to secondaries our picture evolves into the "classical" reacceleration
model (eg., Brunetti et al 2001, 2004; Petrosian 2001; Cassano \& Brunetti 2005) 
where indeed the luminosity of giant radio halos is suppressed
by several orders of magnitude when galaxy clusters evolve into relaxed systems
(e.g. Brunetti et al 2009b; Cassano 2010).

\noindent
Consequently, we believe that
future gamma ray and radio observations will be crucial also
to constrain the ratio of primary and secondary electrons in the IGM.

\section{Acknowledgments}
We thank the anonymous referee for useful comments.
We acknowledge grants from INAF (PRIN-INAF2007 and 2008), ASI-INAF (I/088/06/0), 
NSF (AST 0808118) and NASA (NNX09AH78G) and support by the
NSF Center for Magnetic Self-Organization.
GB thanks the Dep. of Astronomy of Wisconsin University at Madison
and the Harvard-Smithsonian Center for Astrophysics for hospitality.

\section{Appendix: Magnetic field reconnection as source of primary electrons in the
IGM}

Magnetic reconnection is an ubiquitous process in magnetized flows, 
expected to happen when magnetic fields of non-parallel direction get into contact. 
However, the textbook processes of 
magnetic reconnection seem to fall short of providing the desired solution.
The Sweet-Parker reconnection (Sweet 1958, Parker 1957) is too slow.
The speed of the 
reconnection scales as Alfven velocity $v_A$ times the inverse value of 
the square root of the Lunquist number $S^{1/2} \equiv (Lv_A/\eta)^{1/2}$, 
where $L$ is the size of the magnetic regions and $\eta$ is magnetic 
diffusivity. As Lunquist number can be for the IGM $10^{15}$ and higher, 
it is clear that the Sweet-Parker reconnection can handle only a negligible 
fraction of magnetic flux in the Hubble time. 
Petscheck proposed a solution where magnetic fields get into contact at 
a sharp angle (Petscheck 1964).        

\begin{figure}
\includegraphics[width=\columnwidth]{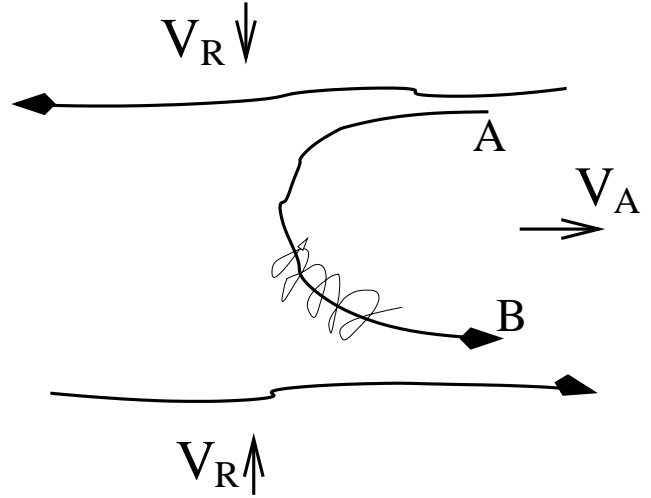}%
\caption{Cosmic rays entrained over a shrinking loop of reconnected magnetic flux
bounce between points A and B and gain energy (from Lazarian 2005).}
\label{fig:accel}
\end{figure}
         
In the Petscheck model no efficient acceleration of particles is expected at 
reconnection sites.
Indeed the traditional processes of acceleration which rely on the electric field 
in the reconnection region are inefficient, as the reconnection region is 
too small and releases only a small portion of magnetic energy. 
Slow shocks predicted in the Petscheck model are likely to be inefficient 
for the particle acceleration (see discussion in Beresnyak, 
Jones \& Lazarian 2009). In addition, observational data does not support the X-point 
reconnection predicted in the Petscheck model either (see Ciaravella \& Raymond 2008). 

A shortcoming of many discussions of magnetic reconnection is that the 
traditional setup does not include ubiquitous pre-existing astrophysical 
turbulence. 
As turbulence radically changes many astrophysical processes, 
the influence of turbulence on reconnection has attracted the attention of 
researchers for a long time (see Speizer 1970;
Mathaeus \& Lamkin 1985, 1986; Strauss 1988).

\noindent
A new approach to the effects of turbulence was adopted 
in Lazarian \& Vishniac (1999, henceforth LV99). 
This model predicts reconnection speeds close to the turbulent velocity in 
the fluid. More precisely, assuming isotropically driven turbulence 
characterized by an injection scale, $l$, smaller than the current sheet 
length $L$, LV99 obtained :

\begin{equation}
V_{rec}\approx v_A\left(l/L\right)^{1/2}\left(V_l/v_A\right)^2,
\label{eq:recon0}
\end{equation}

where the turbulent injection velocity $V_l$ is assumed to be less than $v_A$. 
If $L<l$, the first factor in Eq.~(\ref{eq:recon0}) should be changed to 
$(L/l)^{1/2}$ (LV99). 
If turbulent injection velocity is larger than $v_A$ 
the reconnection happens at the Alfven speed for $L>l_A$. For $L<l_A$ a factor
$(L/l_A)^{1/2}$ should substitute the factor of $(l/L)^{1/2}$ in 
Eq.~(\ref{eq:recon0}). Physically this reflects the fact that at sufficiently 
small scales magnetic field energy dominate the kinetic energy 
and the magnetic 
field lines get only weakly perturbed by turbulent motions. 
Figure~\ref{fig:accel} provides the simplest realization of the acceleration
of particles within the reconnection region expected within LV99 model. 
As a particle bounces back and forth between converging magnetic fluxes, 
it gains energy through the first order Fermi acceleration 
(de Gouveia dal Pino \& Lazarian 2003, 2005; see also Lazarian 2005).
The first order acceleration of particles entrained on contracting magnetic 
loop can be understood from the Liouville theorem, i.e the preservation of 
the phase volume which includes the spatial and momentum coordinates. 
As in the process of reconnection the magnetic tubes are contracting and 
the configuration space presented by magnetic field shrinks, the regular 
increase of the particle's energies is expected. 
The requirement for the process to proceed efficiently is to keep the 
accelerated particles within the contracting magnetic loop. 
This introduces limitations on the particle diffusivities perpendicular 
to magnetic field direction.
Thus high perpendicular diffusion of particles may decouple them from the 
magnetic field. Indeed, it is easy to see that while the particles within 
a magnetic flux rope depicted in Figure~\ref{fig:accel} bounce back and forth 
between the converging mirrors and get accelerated, if these particles 
leave the flux rope fast, they may start bouncing between the magnetic 
fields of different flux ropes which may sometimes decrease their energy. 
Thus it is important that the particle diffusion parallel and perpendicular 
magnetic field stays different. Particle anisotropy which arises from 
particle preferentially getting acceleration in terms of the parallel 
momentum may also be important.
The energy spectrum was derived in GL03 :

\begin{equation}
N(E)dE=C\cdot E^{-5/2}dE,
\label{-5/2}
\end{equation}

In Drake et al. (2006) this idea was enriched by taking into account the 
backreaction of particles. 

\noindent
We believe that such an acceleration can be present in IGM.
Two dimensional numerical simulations of acceleration can be found in  
Drake et al. (2006, 2010) and first three dimensional simulations were 
presented in Lazarian et al. (2010). In typical IGM the small scale 
reconnection of turbulent field should be collisionless and during this 
collisionless with electrons and ions decoupled. 
This should enable the acceleration of electrons.
The idea of particle acceleration in reconnection regions was recently 
applied to describe anomalous cosmic ray acceleration in Heliosphere 
(Lazarian \& Opher 2009, Drake et al.2010) and acceleration of cosmic rays 
in heliotail (Lazarian \& Desiati 2010).
In IGM magnetic turbulence creates magnetic reversals and therefore we 
expect to see additional sources of energetic electrons. 
In terms of the model that we discuss the electron acceleration in 
magnetic reconnection sites should increase the energy density of 
electrons and modify their spectrum. 
We shall discuss these effects elsewhere.

\end{document}